\documentclass[preprint2]{aastex631}

\shorttitle{NEOWISE Thermophysical Modeling of Apophis}
\shortauthors{Satpathy et al.}

\graphicspath{{./}{figs/}}

\begin{document}

\title{\Large{NEOWISE Observations Of The Potentially \\Hazardous Asteroid (99942) Apophis}}

\author[0000-0001-5766-8819]{Akash Satpathy}
\affiliation{Lunar and Planetary Laboratory, University of Arizona, Tucson, AZ, USA}
\email{satpathyakash@email.arizona.edu}

\author[0000-0002-7578-3885]{Amy Mainzer}
\affiliation{Lunar and Planetary Laboratory, University of Arizona, Tucson, AZ, USA}

\author[0000-0003-2638-720X]{Joseph R. Masiero}
\affiliation{Caltech-IPAC, Pasadena, CA, USA}

\author[0000-0003-2534-673X]{Tyler Linder}
\affiliation{Astronomical Research Institute, Ashmore, IL, USA}

\author[0000-0002-0077-2305]{Roc M. Cutri}
\affiliation{Caltech-IPAC, Pasadena, CA, USA}

\author[0000-0001-5058-1593]{Edward L. Wright}
\affiliation{UCLA Astronomy, Los Angeles, CA, USA}

\author{Jana Pittichová}
\affiliation{Jet Propulsion Laboratory, California Institute of Technology, Pasadena, CA, USA}

\author{Tommy Grav}
\affiliation{Lunar and Planetary Laboratory, University of Arizona, Tucson, AZ, USA}

\author[0000-0003-0457-2519]{Emily Kramer}
\affiliation{Jet Propulsion Laboratory, California Institute of Technology, Pasadena, CA, USA}

\DeclareRobustCommand{\rchi}{{\mathpalette\irchi\relax}}
\newcommand{\irchi}[2]{\raisebox{\depth}{$#1\chi$}} 

\begin{abstract}
Large potentially hazardous asteroids (PHAs) are capable of causing a global catastrophe in the event of a planetary collision. Thus, rapid assessment of such an object's physical characteristics is crucial for determining its potential risk scale. We treated the near-Earth asteroid (99942) Apophis as a newly discovered object during its 2020-2021 close-approach as part of a mock planetary defense exercise. The object was detected by the Near-Earth Object Wide-field Infrared Survey Explorer (NEOWISE), and data collected by the two active bands (3.4 {\textmu}m and 4.6 {\textmu}m) were analyzed using thermal and thermophysical modeling. Our results indicate that Apophis is an elongated object with an effective spherical diameter D\textsubscript{eff} = 340 $\pm$ 70 m, a geometric visual albedo p\textsubscript{V} = 0.31 $\pm$ 0.09, and a thermal inertia $\Gamma$ $\sim$ 150 - 2850 Jm{\textsuperscript{-2}}s{\textsuperscript{-{\onehalf}}}K{\textsuperscript{-1}} with a best-fit value of 550 Jm{\textsuperscript{-2}}s{\textsuperscript{-{\onehalf}}}K{\textsuperscript{-1}}. NEOWISE ``discovery" observations reveal that (99942) Apophis is a potentially hazardous asteroid that would likely cause damage at a regional level and not a global one.
\end{abstract}


\section{\textbf{Overview}} \label{sec:intro}

Near-Earth Asteroids (NEAs) are a sub-population of asteroids that pass very close to the Earth. They are conventionally defined as small bodies with a perihelion distance less than or equal to 1.3 au, and compared to Main-Belt Asteroids (MBAs), they tend to be smaller in size and irregularly shaped with short dynamical lifespans. Due to their relative proximity to the Earth, detailed studies of their physical characteristics such as diameter, albedo, thermal inertia, rotational period, and absolute visual magnitude can be performed by using data obtained from either remote sensing (Ostro \citeyear{Ostro_radar}, Werner et al. \citeyear{Spitzer}, and Mainzer et al. \citeyear{mainzer_prelimresults_11}) or in-situ (Watanabe et al. \citeyear{hayabusa2}, Lauretta et al. \citeyear{orex}) missions. Studying these features can help us theorize their origin and evolution in the solar system (as laid out by Binzel et al. \citeyear{Binzel_origin}, Michel et al. \citeyear{NEO_origin_Bottke}, Granvik \& Brown \citeyear{gravnik_meteor_origin}, and others), and take pivotal steps towards planetary defense.\\

NEAs with a Minimum Orbit Intersection Distance \href{http://www2.lowell.edu/users/elgb/moid.html}{(MOID)} to the Earth of less than 0.05 au and an H magnitude less than or equal to 22 (or correspondingly, a minimum diameter of roughly 140 meters) are formally classified as Potentially Hazardous Asteroids or PHAs as they have the capability to cause substantial damage upon collision with the Earth. One example from recorded history is the 1908 Tunguska event, where a meteor with a diameter of $\sim$100 meters exploded over a Siberian forest, producing around 60 PJ of energy and affecting an area as large as 2150 km\textsuperscript{2} (Vasilyev \citeyear{tunguska}). A more recent and better-documented example is the 2013 Chelyabinsk event where a small NEA about 19 meters in diameter (Borovi{\v{c}}ka et al. \citeyear{chelyabinsk}) exploded over Chelyabinsk Oblast, Russia, injuring roughly 1500 people and damaging around 7200 buildings. Therefore, the discovery and categorization of such asteroids are crucial to prevent potential impacts and mitigate harm.\\

A particular NEA --- (99942) Apophis --- has been an object of great interest since its discovery in 2004 by R. A. Tucker, D. J. Tholen, and F. Bernardi. It is a well-known PHA with a catalog H magnitude of 19.7, a perihelion distance of roughly 0.746 au, and a MOID of 0.0002056 au. Thousands of optical and dozens of radar observations have allowed planetary scientists to study Apophis in great detail. Binzel et al. \citeyearpar{binzel_Sq} studied its spectral properties and composition and categorized it as an Sq-class asteroid (Bus \citeyear{bus_sq} and DeMeo et al. \citeyear{demeo_bus_extend}), and Lin et al. \citeyearpar{Lin_S} classified it as an S-class asteroid (Tholen \citeyear{tholen_S}) after carrying out a photometric survey. Delbo et al. \citeyearpar{delbo} used polarimetric observations to estimate its size and albedo, and Müeller et al. \citeyearpar{mueller} used thermal measurements from Herschel to determine its size, albedo, thermal inertia, and also calculated its mass by using Itokawa's density and porosity. The most recent results based on radar observations from Goldstone and Arecibo suggest that Apophis is an elongated, asymmetric, and possibly bifurcated object with a diameter of 340 $\pm$ 40 m and a visible geometric albedo of 0.35 $\pm$ 0.10 (Brozovi\'{c} et al. \citeyear{brozovic}).\\

Post-discovery, Apophis drew international attention when its collision probability was initially estimated to be as high as 2.7\% for the year 2029. It was speculated that it might enter a so-called gravitational keyhole, which could exacerbate the possibility of an impact in the year 2036. Follow-up studies and analysis significantly lowered this probability, and any chance of collision in 2029 and 2036 was ruled out. Recent radar observations from Goldstone also eliminated the slight risk of impact for the year 2068 and further showed that it posed no threat to the Earth for at least another hundred years.\footnote{\url{https://echo.jpl.nasa.gov/asteroids/Apophis/apophis.2021.goldstone.planning.html}} Nonetheless, Apophis remained a "virtual impactor" of special interest to the planetary defense community and was the subject of a mock planetary defense exercise during its 2020-2021 flyby. \\

In the exercise, Apophis was treated as a newly discovered asteroid, and the capability of research groups to identify and rapidly characterize potentially hazardous objects were tested (inspired by a similar activity performed by Reddy et al. \citeyear{reddy_pdefense} where they tracked and characterized the NEA 2012 TC4 as a hypothetical impactor). Thermal infrared data on (99942) Apophis was collected by the Near-Earth Object Wide-field Infrared Survey Explorer (NEOWISE; Mainzer et al. \citeyear{mainzer_reactivation}, Wright et al. \citeyear{wright_10}) during the object’s close approaches in December of 2020 and March-April of 2021. These new observations resulted in the “discovery” of the object, and its diameter and albedo were rapidly computed following standard image processing and calibration by the NEOWISE data system (Cutri et al. \citeyear{cutri_neowise}). While a forthcoming publication will describe the full results of the mock exercise, our study reconfirms prior results and highlights the accuracy and speed of the NEOWISE team’s analytical methods.

\tabcolsep=0.1cm
\begin{deluxetable*}{clcccc}
\tablenum{1}
\tablecaption{Observing geometry of (99942) Apophis during the two NEOWISE observing epochs.\label{tab:apophis2}}
\tablewidth{0pt}
\tablehead{
\colhead{Epoch} & \colhead{MJD} & \colhead{WISE-centric distance} & \colhead{Heliocentric distance} & \colhead{Solar Elongation} & \colhead{Phase Angle}\\
\nocolhead{} & \nocolhead{} & \colhead{\footnotesize{au}} & \colhead{\footnotesize{au}} & \colhead{\footnotesize{Degrees}} & \colhead{\footnotesize{Degrees}} 
}
\startdata
1 & 59202.561219 & 0.259 & 1.018 & 90.2 & 75.1\\
2 & 59305.273533 & 0.137 & 1.057 & 111.2 & 61.8\\
\enddata
\tablecomments{For Epoch 1, the start and end date and time were 18-12-2020 at 23:26:13 and 20-12-2020 at 08:23:39, respectively. For Epoch 2, the start and end date and time were 31-03-2021 at 21:49:28 and 01-04-2021 at 15:04:56, respectively.\label{fig:table_small}}
\end{deluxetable*}

\tabcolsep=0.1cm
\begin{deluxetable*}{clccccc}
\tablenum{2}
\tablecaption{Truncated table of CTIO observations of (99942) Apophis in B, \textit{g}, \textit{r}, and V.\label{tab:v_table}}
\tablehead{
\colhead{Filter} & \colhead{MJD} & \colhead{RA} & \colhead{Dec} & \colhead{Zero Point} & \colhead{Instantaneous Mag} & \colhead{Calibrated Mag}\\ 
\nocolhead{} & \nocolhead{} & \colhead{\footnotesize{Degrees}} & \colhead{\footnotesize{Degrees}} & \colhead{\footnotesize{Mag}} & \colhead{\footnotesize{Mag}} & \colhead{\footnotesize{Mag}}
}
\startdata
 & 59201.290444 & 172.42233 & -10.59855 & 27.342 ± 0.048 & -7.630 ± 0.133 & 19.713 ± 0.141\\
\textbf{B} & 59201.294088 & 172.42300 & -10.59945 & 27.575 ± 0.057 & -7.995 ± 0.104 & 19.581 ± 0.118\\
 & 59201.297726 & 172.42366 & -10.60035 & 27.382 ± 0.061 & -7.632 ± 0.122 & 19.750 ± 0.136\\
 & & & & \textbf{...}\\
 & 59201.291203 & 172.42246 & -10.59873 & 26.997 ± 0.038 & -8.319 ± 0.066 & 18.678 ± 0.077\\
 & 59201.294848 & 172.42313 & -10.59963 & 27.010 ± 0.053 & -8.210 ± 0.070 & 18.800 ± 0.088\\
 & 59201.298486 & 172.42379 & -10.60053 & 27.012 ± 0.049 & -8.223 ± 0.062 & 18.789 ± 0.079\\
 & 59201.302166 & 172.42459 & -10.60161 & 27.022 ± 0.037 & -8.386 ± 0.058 & 18.636 ± 0.069\\
 & 59201.305797 & 172.42525 & -10.60251 & 27.029 ± 0.044 & -8.435 ± 0.065 & 18.594 ± 0.078\\
 & 59201.309434 & 172.42590 & -10.60341 & 27.050 ± 0.053 & -8.517 ± 0.079 & 18.533 ± 0.095\\
 & 59201.313071 & 172.42655 & -10.60431 & 27.041 ± 0.060 & -8.616 ± 0.065 & 18.425 ± 0.089\\
 & 59201.316718 & 172.42733 & -10.60539 & 27.058 ± 0.036 & -8.665 ± 0.059 & 18.393 ± 0.070\\
\textbf{V} & 59201.320353 & 172.42798 & -10.60629 & 27.057 ± 0.036 & -8.520 ± 0.066 & 18.538 ± 0.075\\
 & 59201.323983 & 172.42862 & -10.60719 & 27.075 ± 0.055 & -8.434 ± 0.056 & 18.641 ± 0.078\\
 & 59201.327613 & 172.42927 & -10.60809 & 27.071 ± 0.050 & -8.379 ± 0.065 & 18.692 ± 0.082\\
 & 59201.331241 & 172.42990 & -10.60899 & 27.045 ± 0.057 & -8.479 ± 0.054 & 18.566 ± 0.078\\
 & 59201.334881 & 172.43067 & -10.61006 & 27.029 ± 0.047 & -8.484 ± 0.057 & 18.545 ± 0.074\\
 & 59201.338529 & 172.43130 & -10.61096 & 27.056 ± 0.044 & -8.447 ± 0.053 & 18.608 ± 0.069\\
 & 59201.342191 & 172.43193 & -10.61185 & 27.022 ± 0.053 & -8.296 ± 0.068 & 18.726 ± 0.086\\
 & 59201.345835 & 172.43269 & -10.61293 & 27.045 ± 0.047 & -8.543 ± 0.052 & 18.502 ± 0.070\\
\enddata
\tablecomments{The CTIO V-filter observations began on 18-12-2020 at 06:59:20 and ended on 18-12-2020 at 08:18:00 (i.e., separated from NEOWISE's first epoch of observations by roughly 15 hours). The airmass during the first image was 1.6, and the same during last image was 1.22. Airmass was reducing during the observing run. All observations were taken during night-time and were four days past new moon (i.e., there was no moon effect). The table has been truncated for conciseness. All the values below the ellipses are in V, and observations in B, \textit{g}, and \textit{r} filters can be accessed in an electronic format. Only the median V-magnitude (18.60 $\pm$ 0.11) was used in this paper to constrain the H-mag.
\label{fig:v_table}}
\end{deluxetable*}

\section{\textbf{Observations}} \label{sec:obs}

The Near-Earth Object Wide-field Infrared Survey Explorer (NEOWISE; Mainzer et al. \citeyear{mainzer_reactivation}) is a two-band all-sky thermal infrared survey well suited to investigating the physical properties of asteroids and comets. It was initially launched as the Wide-field Infrared Survey Explorer (WISE; Wright et al. \citeyear{wright_10}) in December of 2009 with the objective to map 99\% of the sky using four thermal IR bands with wavelengths 3.4, 4.6, 12, and 22 {\textmu}m (named W1, W2, W3, and W4, respectively). The primary mission concluded after the solid hydrogen coolant was exhausted and the W3 and W4 channels ceased to function, and WISE was put into hibernation in February 2011.  In December 2013, the spacecraft was reactivated and repurposed as NEOWISE with the W1 and W2 channels active, and since that time has detected over 40,000 different solar system small bodies.\footnote{\url{https://wise2.ipac.caltech.edu/docs/release/neowise/}}\\

\begin{figure}[bht!]
\begin{interactive}{animation}{apophis_flyby_20_21.mp4}
\plotone{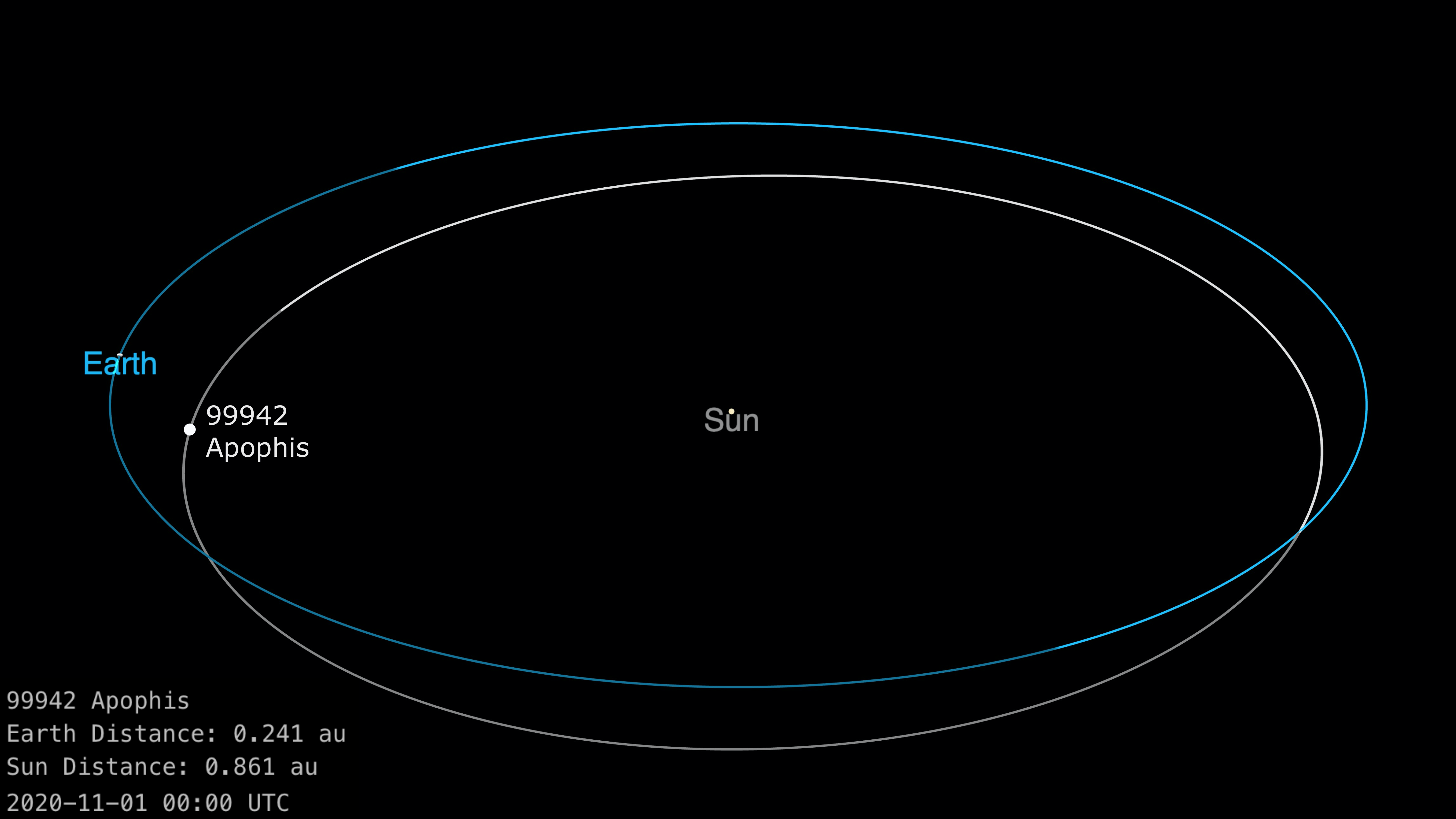}
\end{interactive}
\caption{A seventeen second long video that shows (99942) Apophis and the Earth's motion from November 2020 to May 2021. The animation is available in the HTML version of this article. The first and second epoch of observations are labelled with the help of freeze frames. The name of object, the distance to Earth and the Sun, and the date and time are notated at the bottom-left corner. The video was made with the help of the \href{https://ssd.jpl.nasa.gov/sbdb.cgi}{JPL Small-Body Database Browser}.\label{fig:video}}
\end{figure}

NEOWISE was one of the first surveys to submit “discovery” tracklets (sets of position-time measurements of candidate moving objects) for Apophis late in 2020. Thirty-one single exposure detections were acquired during the two observing epochs (see figure \ref{fig:video} for reference). First, the detections were automatically ingested by the WISE Moving Object Pipeline Subsystem (WMOPS) that associates sets of detections with apparent on-sky motions that are consistent with orbital motion. Detections were rejected if they had moon separation angles (\textit{moon\_sep}) of less than 15$^{\circ}$, were saturated,\footnote{\url{https://wise2.ipac.caltech.edu/docs/release/neowise/expsup/sec2\textunderscore1civa.html}} were co-located within 6.5 arcsec of stationary, background objects such as stars or galaxies (identified from the AllWISE Atlas Image set from Cutri et al. \citeyear{cutri_allwise}), or had poor fits to the reference point spread functions i.e. a \textit{w1rchi2} $\geq$ 5 and \textit{w2rchi2} $\geq$ 5 (most likely due to cosmic rays).\footnote{\url{https://wise2.ipac.caltech.edu/docs/release/neowise/expsup/sec2_1a.html}} Lastly, the detection images were visually inspected to check for any other unflagged artifacts, and additional scan frames were extracted using the WISE Moving Object Search Tool (MOST) as a few detections were rejected by WMOPS due to the curvature of their on-sky motion or falling near a background object.\footnote{\url{https://irsa.ipac.caltech.edu/applications/MOST/}}\\

\begin{figure*}
\plotone{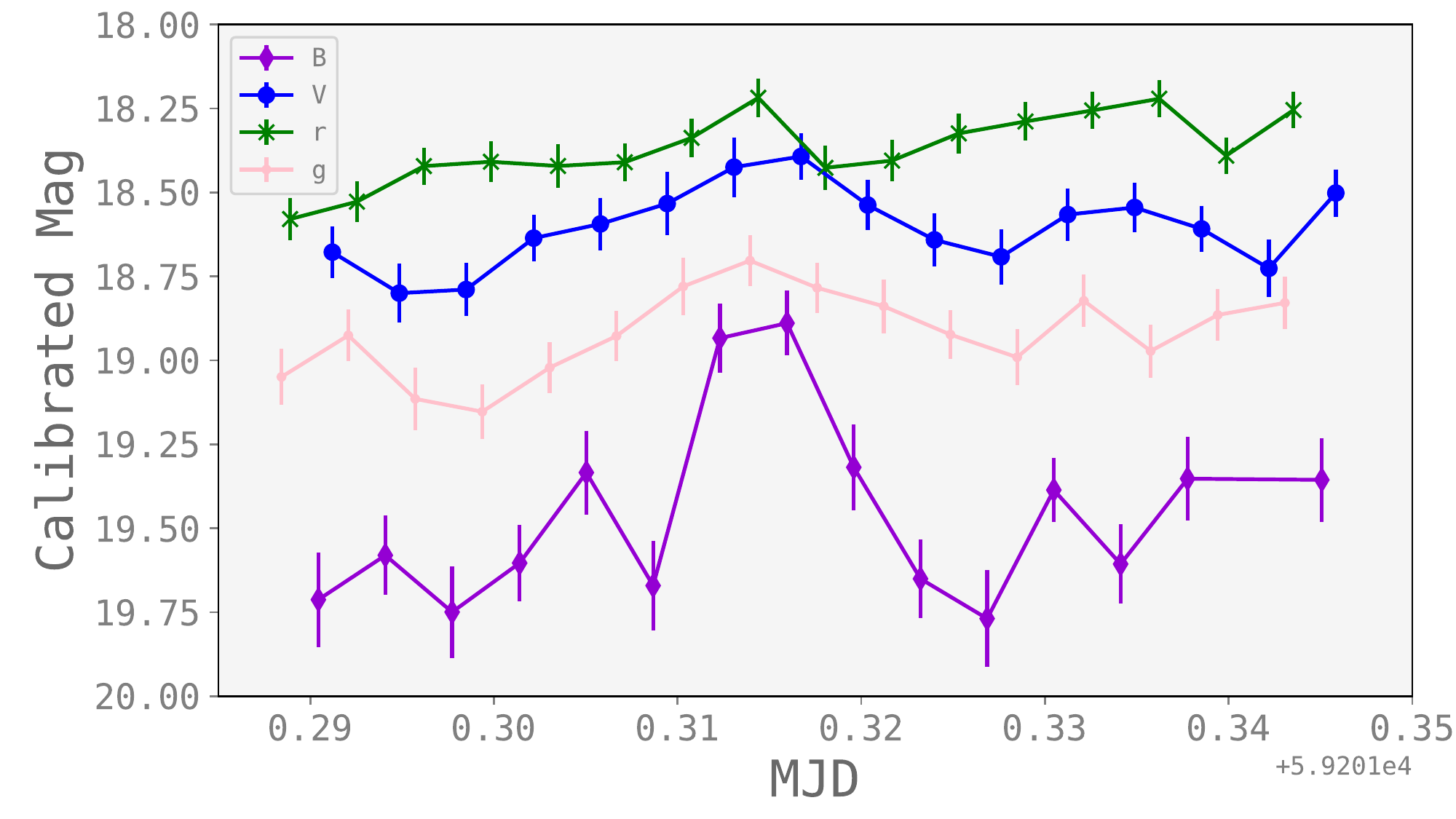}
\caption{CTIO observations in B, V, \textit{r}, and \textit{g}; taken at nearly the same time as NEOWISE's first epoch of observations.
\label{fig:Vmag_CTIO}}
\end{figure*}

From the first epoch in December 2020, seventeen detections were identified in the two active bands (3.4 {\textmu}m and 4.6 {\textmu}m), with an observation time spanning 32.75 hours. The SNR at 4.6 {\textmu}m was approximately 5.5, and the solar elongation angle was roughly 90$^{\circ}$. Eight additional detections were collected during the second observing epoch in April 2021, spanning 17.25 hours with an average SNR of 32.7 at 4.6 {\textmu}m and a solar elongation angle of about 110$^{\circ}$. The SNR was noticeably higher during the second epoch because Apophis was closer to the Earth than it was during the first epoch. A summary of the observing geometry can be found in Table \ref{tab:apophis2}, and detailed information about the same is given in Table \ref{fig:table_big}.\\

Lastly, visible photometry nearly simultaneous to Epoch 1 was obtained using the SMARTS 1.0-meter telescope from the Cerro Tololo Inter-American Observatory (CTIO) was used to provide an improved constraint on the absolute visual magnitude i.e. H (as H magnitudes for many asteroids, particularly near-Earth objects, often have large uncertainties, which in turn reduces the accuracy of derived visible albedos). SMARTS 1.0 meter is a Boller and Chivens f/10.5 telescope with a back-illuminated Finger Lakes Instruments thermo-electrically cooled camera that has 13.5 {\textmu}m pixels and an array size of 2048 x 2048 pixels. Data were taken continuously on the 18th of December, 2020 from 06:54 to 08:18 UTC with filters in the following order: Clear, \textit{g}, \textit{r}, \textit{i}, \textit{z}, B, and, V. There was a $\sim$3 second delay between exposures while the filter wheel moved.  In total, 112 images were taken, i.e., 16 exposures for each of the seven filters. The pixel binning size was set to 4, which produced a 1.05 arcsec per pixel image scale. The data were calibrated to Pan-STARRS, and Kostov \& Bonev \citeyearpar{kostov_transformation} were referenced to convert Pan-STARRS \textit{g}, \textit{r}, \textit{i} values to B and V (see Table 2 in that paper). These transformed V values, listed in Table \ref{tab:v_table} and plotted in Figure \ref{fig:Vmag_CTIO}, were then used to calibrate the V filter and compute an H = 19.1 $\pm$ 0.5 using the $HG$-system delineated in Bowell et al. \citeyear{bowell_89} (G = 0.25 $\pm$ 0.20 from Shevchenko et al. \citeyear{shevchenko_phase} was assumed during the conversion).

\tabcolsep=0.1cm
\begin{deluxetable*}{llllccccl|ccccl}
\tablenum{3}
\tablecaption{Two-epoch observations of (99942) Apophis in the 3.4 {\textmu}m and 4.6 {\textmu}m bands.\label{tab:apophis1}}
\tablewidth{1pt}
\tablehead{
\colhead{RA} & \colhead{Dec} & \colhead{MJD} & \colhead{Source ID} & \nocolhead{} & \nocolhead{} & \colhead{W1} & \nocolhead{} & \nocolhead{} & \nocolhead{} & \nocolhead{} & \colhead{W2} & \nocolhead{}& \nocolhead{} \\
\nocolhead{} & \nocolhead{} & \nocolhead{} & \nocolhead{} & \colhead{\scriptsize{Mag}} & \colhead{\scriptsize{Mag $\sigma$}} & \colhead{\scriptsize{Flux}} & \colhead{\scriptsize{Flux $\sigma$}} & \colhead{\scriptsize{$\rchi$\textsuperscript{2}}} & \colhead{\scriptsize{Mag}} & \colhead{\scriptsize{Mag $\sigma$}} & \colhead{\scriptsize{Flux}} & \colhead{\scriptsize{Flux $\sigma$}} & \colhead{\scriptsize{$\rchi$\textsuperscript{2}}} 
}
\decimalcolnumbers
\startdata
172.5647 & -10.7728 & 59201.97654340 & \footnotesize{23614r158-002887} & 16.468 & 0.363 & 52.140 & 17.452 & 0.97 & 14.136 & 0.328 & 159.64 & 48.162 & 0.83\\
172.5647 & -10.7723 & 59201.97667076 & \footnotesize{23614r159-002689} & 15.366 & -- & 66.212 & 38.803 & 0.68 & 14.304 & 0.212 & 136.71 & 26.672 & 1.10\\
172.5938 & -10.8056 & 59202.10731955 & \footnotesize{23618r158-001173} & 16.670 & 0.459 & 43.306 & 18.320 & 0.96 & 13.857 & 0.138 & 206.38 & 26.256 & 0.38\\
172.5941 & -10.8058 & 59202.10744691 & \footnotesize{23618r159-000844} & 15.352 & 0.219 & 145.68 & 29.341 & 0.50 & 14.268 & 0.349 & 141.34 & 45.426 & 1.53\\
172.6229 & -10.8388 & 59202.23809564 & \footnotesize{23622r158-002053} & 16.696 & 0.433 & 42.256 & 16.837 & 1.22 & 14.134 & 0.165 & 159.85 & 24.265 & 0.61\\
172.6232 & -10.8387 & 59202.23822300 & \footnotesize{23622r159-001278} & 16.467 & 0.357 & 52.169 & 17.172 & 0.69 & 13.809 & 0.141 & 215.73 & 27.933 & 0.54\\
172.6808 & -10.9049 & 59202.49926582 & \footnotesize{23630r146-010408} & 16.466 & 0.323 & 52.211 & 15.513 & 2.03 & 14.277 & 0.265 & 140.11 & 34.187 & 0.36\\
172.6957 & -10.9209 & 59202.56509962 & \footnotesize{23632r134-001369} & 15.745 & 0.225 & 101.48 & 21.036 & 0.82 & 14.600 & 0.287 & 104.13 & 27.559 & 0.42\\
172.7095 & -10.9375 & 59202.63042400 & \footnotesize{23634r158-000846} & 15.568 & 0.194 & 119.47 & 21.293 & 1.19 & 13.991 & 0.172 & 182.41 & 28.970 & 0.51\\
172.7097 & -10.9370 & 59202.63055132 & \footnotesize{23634r159-001077} & 15.691 & 0.169 & 106.61 & 16.560 & 1.27 & 13.966 & 0.280 & 186.62 & 48.126 & 0.76\\
172.7240 & -10.9540 & 59202.69587575 & \footnotesize{23636r134-001054} & 16.046 & 0.232 & 76.939 & 16.424 & 1.07 & 13.717 & 0.127 & 234.82 & 27.524 & 1.20\\
172.7384 & -10.9703 & 59202.76120009 & \footnotesize{23638r158-001182} & 15.760 & 0.205 & 100.07 & 18.877 & 0.91 & 13.838 & 0.158 & 210.04 & 30.474 & 0.57\\
172.7531 & -10.9868 & 59202.82665184 & \footnotesize{23640r134-002266} & 16.276 & -- & 30.427 & 15.892 & 0.74 & 14.234 & 0.233 & 145.77 & 31.301 & 0.76\\
172.7674 & -11.0035 & 59202.89197627 & \footnotesize{23642r158-002419} & 16.243 & 0.281 & 64.152 & 16.604 & 0.70 & 14.396 & 0.222 & 125.57 & 25.675 & 0.89\\
172.7817 & -11.0200 & 59202.95742860 & \footnotesize{23644r134-002255} & 16.518 & 0.539 & 49.773 & 24.690 & 0.80 & 14.352 & 0.211 & 130.83 & 25.422 & 0.29\\
172.8100 & -11.0526 & 59203.08820473 & \footnotesize{23648r134-002554} & 16.412 & 0.360 & 54.910 & 18.223 & 0.44 & 14.375 & 0.278 & 128.04 & 32.788 & 0.93\\
172.8671 & -11.1186 & 59203.34975691 & \footnotesize{23656r134-001623} & 16.405 & 0.443 & 55.270 & 22.528 & 1.37 & 13.789 & 0.134 & 219.73 & 27.045 & 0.93\\
\tableline
123.3184 & 10.9412 & 59304.90934705 & \footnotesize{26763r125-000449} & 14.632 & 0.072 & 282.76 & 18.878 & 0.80 & 11.970 & 0.037 & 1173.1 & 40.288 & 0.69\\
123.2764 & 11.0123 & 59305.03999583 & \footnotesize{26767r125-000363} & 14.045 & 0.061 & 485.89 & 27.342 & 1.04 & 11.682 & 0.031 & 1530.2 & 43.602 & 1.38\\
123.2764 & 11.0122 & 59305.04012315 & \footnotesize{26767r126-000374} & 14.126 & 0.055 & 450.68 & 22.704 & 1.36 & 11.793 & 0.042 & 1381.2 & 53.025 & 1.20\\
123.2347 & 11.0830 & 59305.17077185 & \footnotesize{26771r125-000262} & 13.779 & 0.045 & 620.66 & 25.547 & 0.99 & 11.254 & 0.026 & 2269.6 & 53.736 & 0.37\\
123.2140 & 11.1182 & 59305.23622361 & \footnotesize{26773r102-000306} & 13.750 & 0.044 & 637.32 & 26.031 & 3.08 & 11.327 & 0.029 & 2120.6 & 56.527 & 1.21\\
123.1934 & 11.1535 & 59305.30154795 & \footnotesize{26775r125-000328} & 14.081 & 0.052 & 469.71 & 22.316 & 1.45 & 11.529 & 0.031 & 1760.5 & 50.329 & 0.56\\
123.1322 & 11.2589 & 59305.49764841 & \footnotesize{26781r046-000549} & 14.676 & 0.091 & 271.52 & 22.695 & 1.30 & 12.248 & 0.050 & 907.84 & 42.166 & 0.71\\
123.0919 & 11.3287 & 59305.62842442 & \footnotesize{26785r102-000420} & 14.391 & 0.065 & 353.12 & 21.001 & 0.91 & 11.708 & 0.031 & 1493.9 & 42.510 & 0.82\\
\enddata
\tablecomments{The horizontal line near the middle row separates the first and the second set of observations. The units for Ra and Dec are degrees, and that of flux is {\textmu}Jy. Source ID in column (4) is a unique ID comprising of the scan ID, frame number, and source number.``Mag" in columns (5), (6), (10), and (11) refers to the instrumental profile-fit photometry magnitude of the object at the time of the observation. ``$\sigma$" in columns (6), (8), (11), and (13) represents the measurement uncertainty. The detailed methodology for the determination of in-band uncertainties and flux calibration uncertainties can be found in Cutri et al. \citeyearpar{cutri_neowise} Columns (9) and (14) contain the $\rchi^2$ statistic that show the {\href{https://wise2.ipac.caltech.edu/docs/release/neowise/expsup/sec4_2bi.html}{quality of the profile-fit photometry}} of the measurements.\label{fig:table_big}}
\end{deluxetable*}

\begin{figure*}
\gridline{\fig{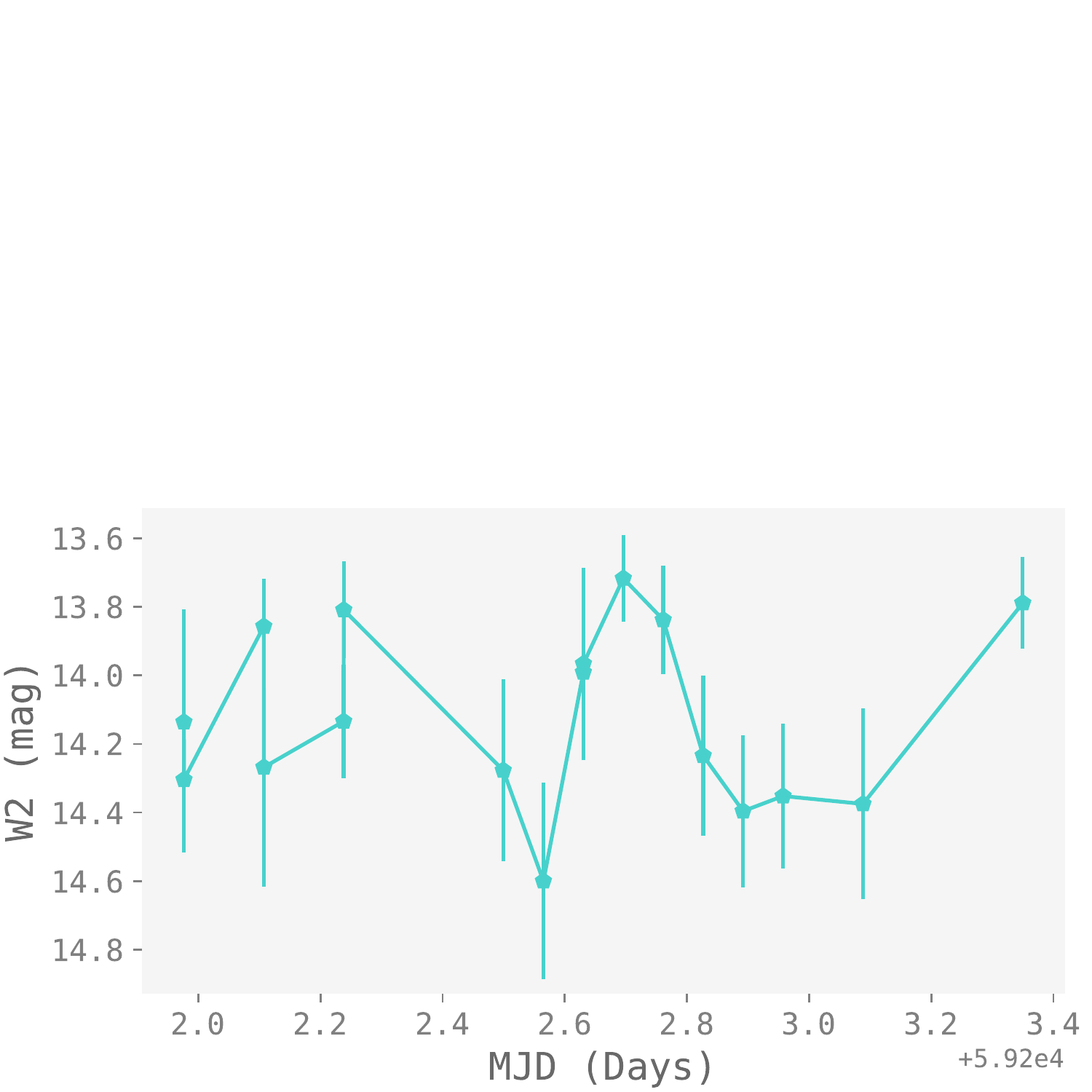}{0.5\textwidth}{}
          \fig{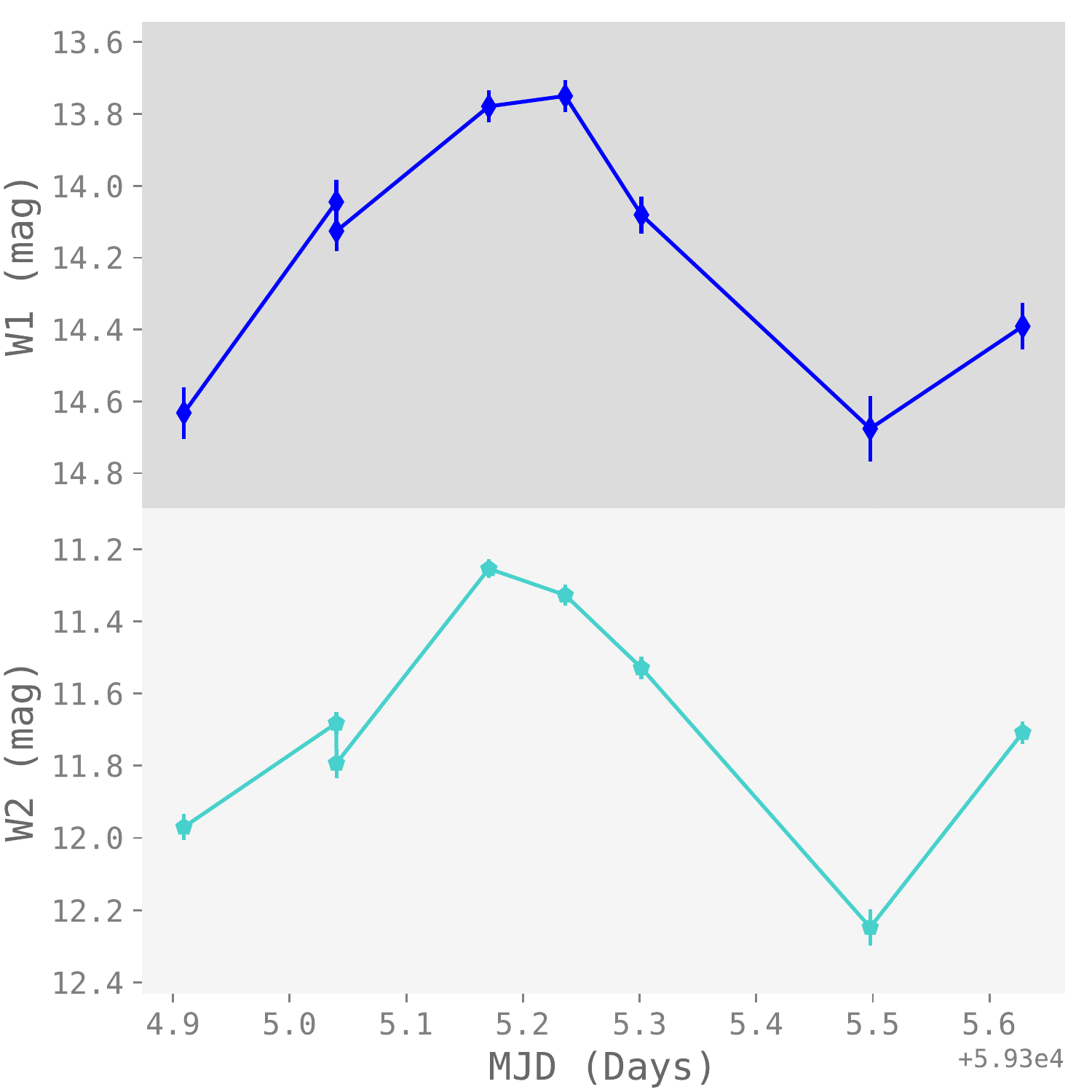}{0.5\textwidth}{}
          }
\caption{Lightcurves from the first and second observing epochs (left and right, respectively). The W1 band data from the first epoch was omitted due to marginal detections. The large amplitude in the lightcurves suggest that the object has an elongated shape, as previously shown by {Brozovi\'{c}} et al. \citeyearpar{brozovic}. 
\label{fig:lightcurves}}
\end{figure*}

\begin{figure*}
\gridline{\fig{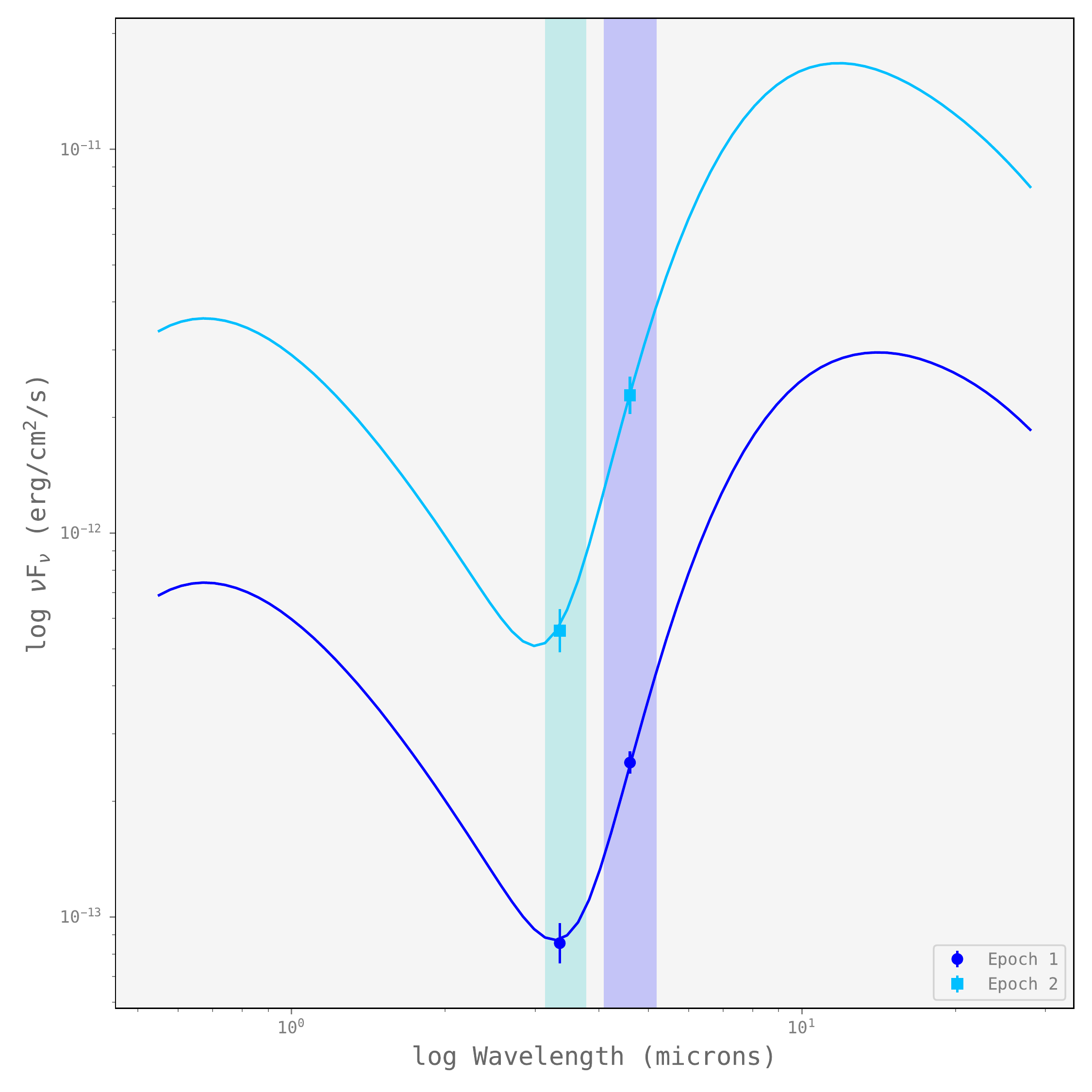}{0.5\textwidth}{}
          \fig{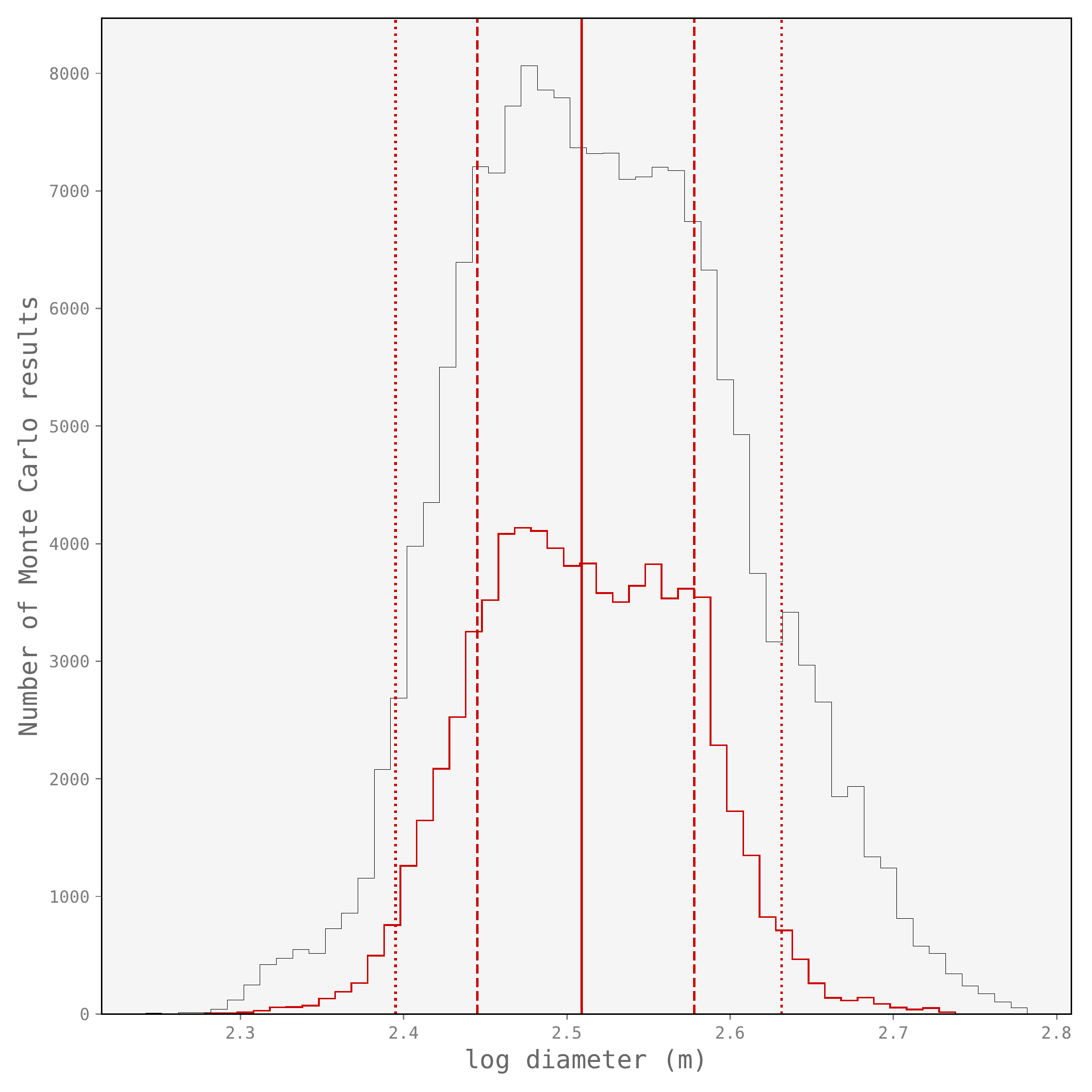}{0.5\textwidth}{}
          }
\gridline{\fig{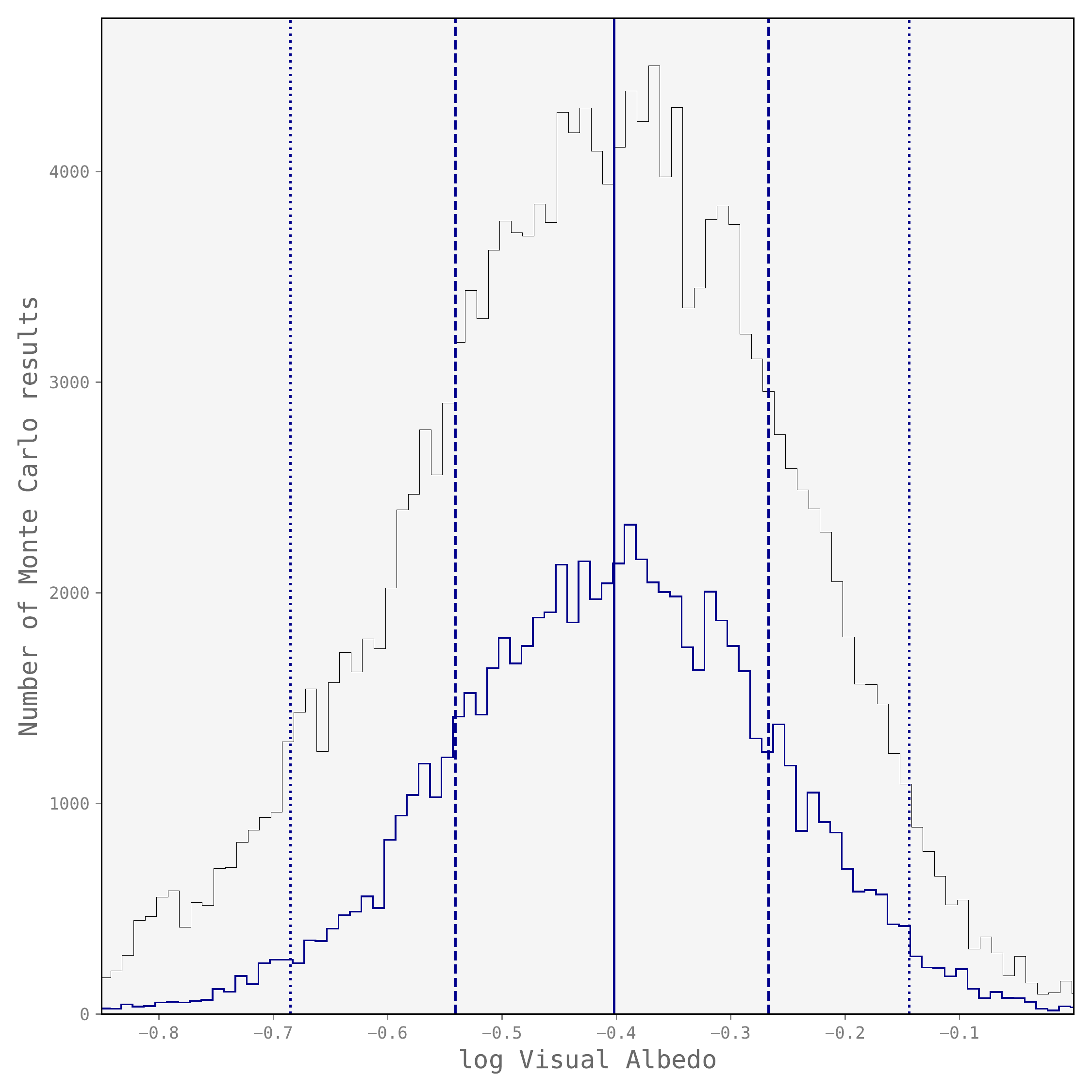}{0.5\textwidth}{}
          \fig{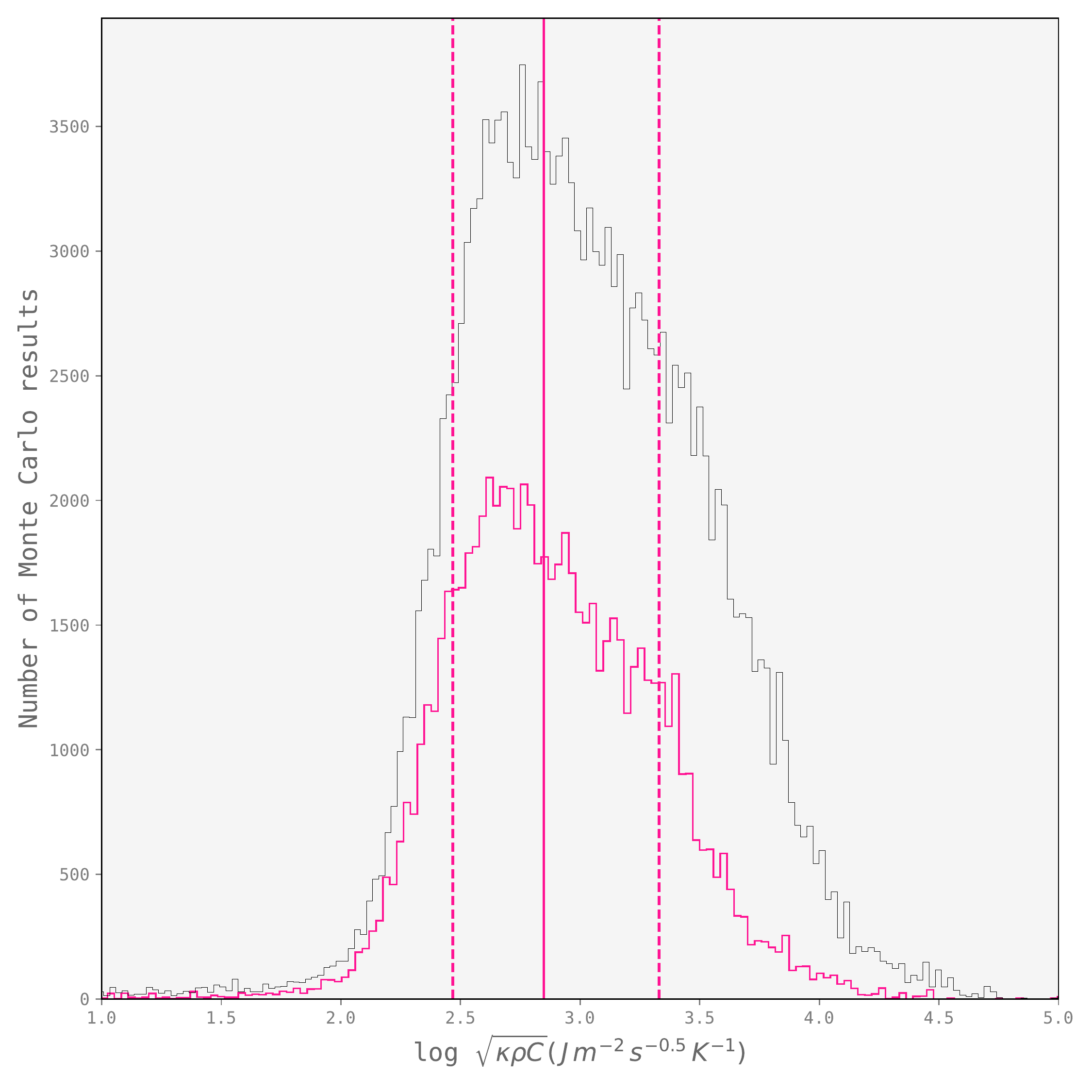}{0.5\textwidth}{}
          }
\caption{Plots highlighting the TPM results with (99942) Apophis modeled as a sphere. In a clockwise manner from the top-left: average flux density during the two epochs along with the span of the bandpasses (the average flux was 95.77 {\textmu}Jy in W1 and 381.13 {\textmu}Jy in W2 during the first epoch, and 622.88 {\textmu}Jy in W1 and 3475.26 {\textmu}Jy in W2 during the second epoch), histogram of the diameter results with the 68\% (dashed lines) and 95\% (dotted lines) confidence interval ranges, histogram of the thermal inertia results with the 68\% (dashed lines) confidence interval range (the 95\% confidence interval values lie outside the x-axis range), and histogram of the visual albedo results with the 68\% (dashed lines) and 95\% (dotted lines) confidence interval ranges. The fainter histograms in the top-right, bottom-left, and bottom-right plots signify the solutions without the LCDB pole solution constraint.\label{fig:TPMplots}}
\end{figure*}

\begin{figure*}
\gridline{\fig{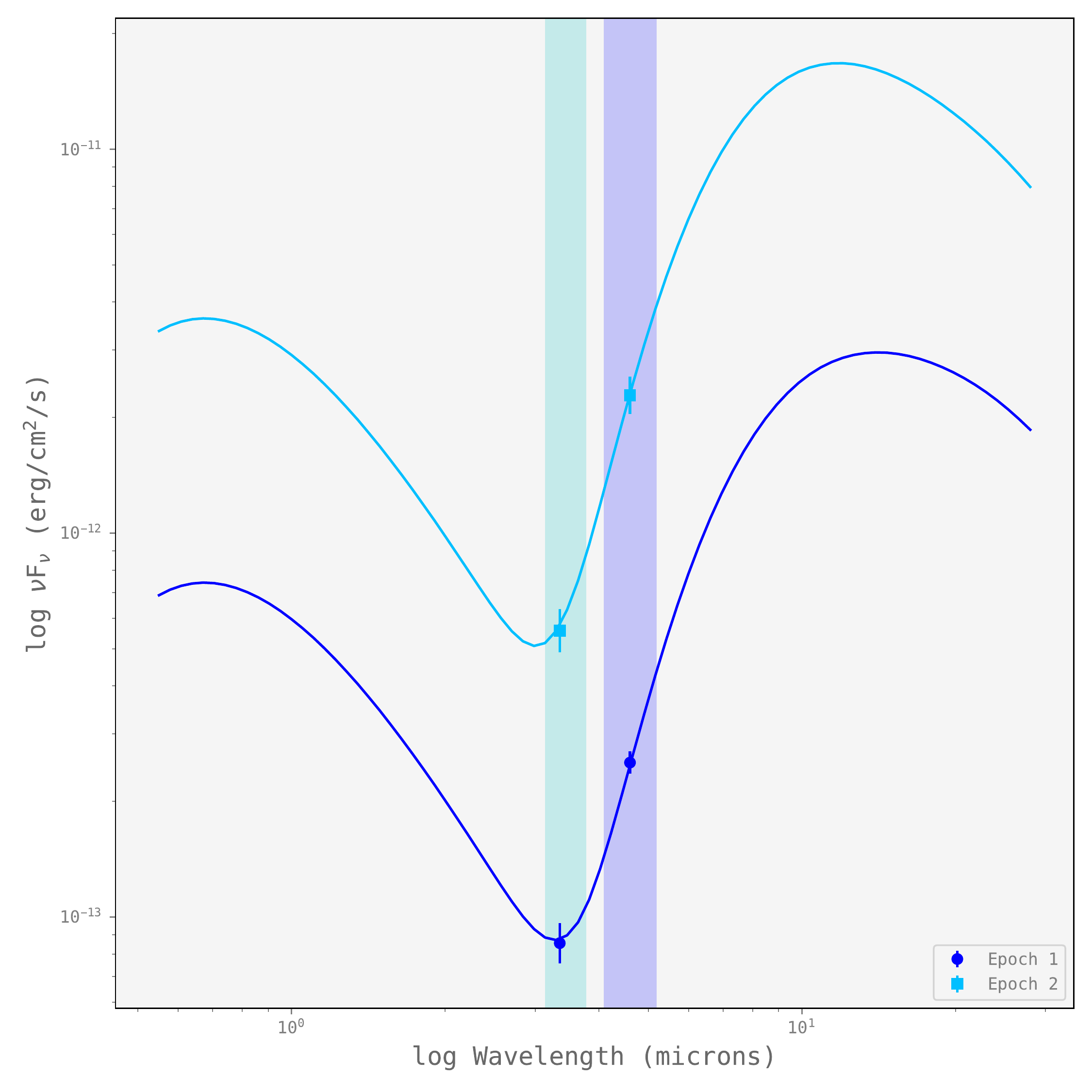}{0.5\textwidth}{}
          \fig{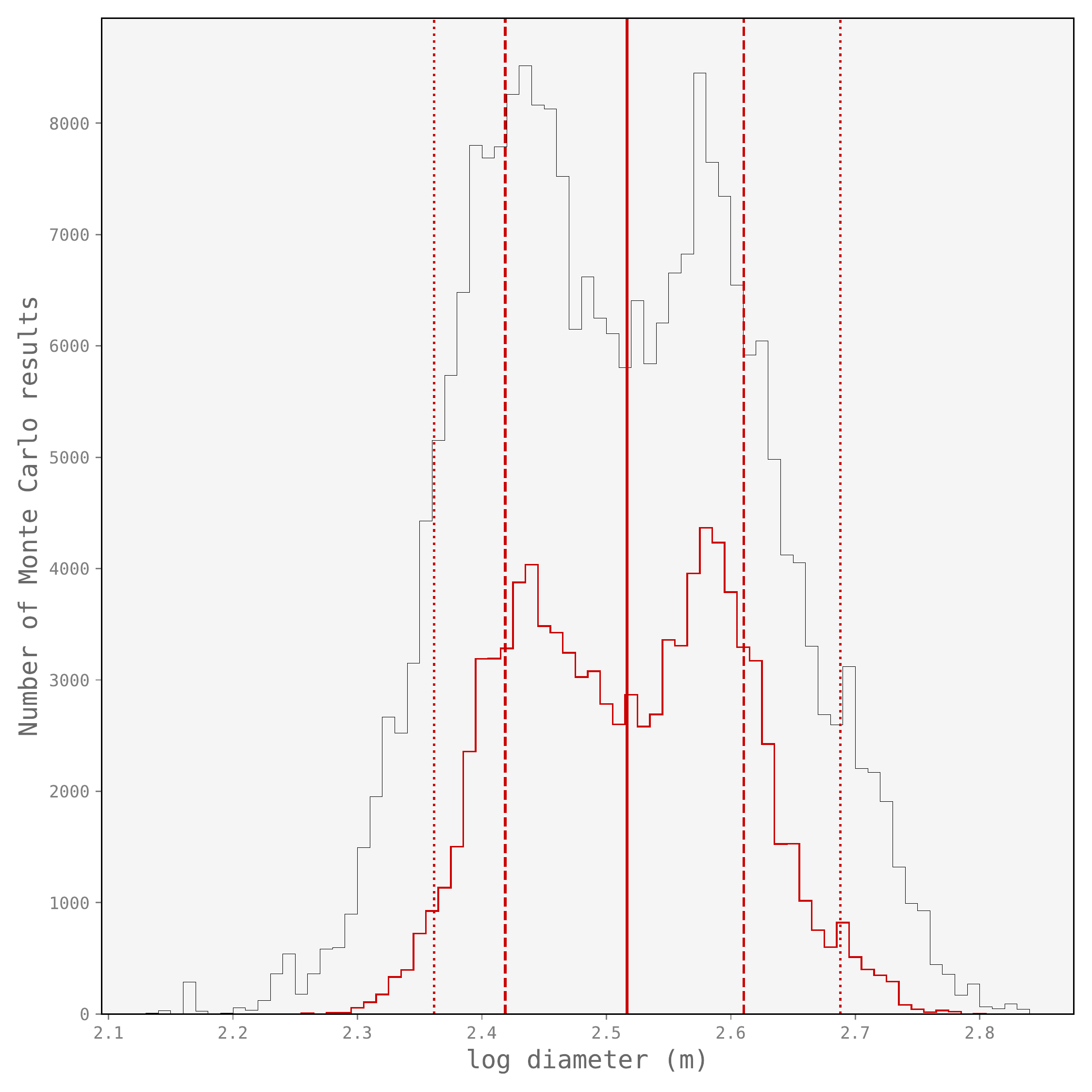}{0.5\textwidth}{}
          }
\gridline{\fig{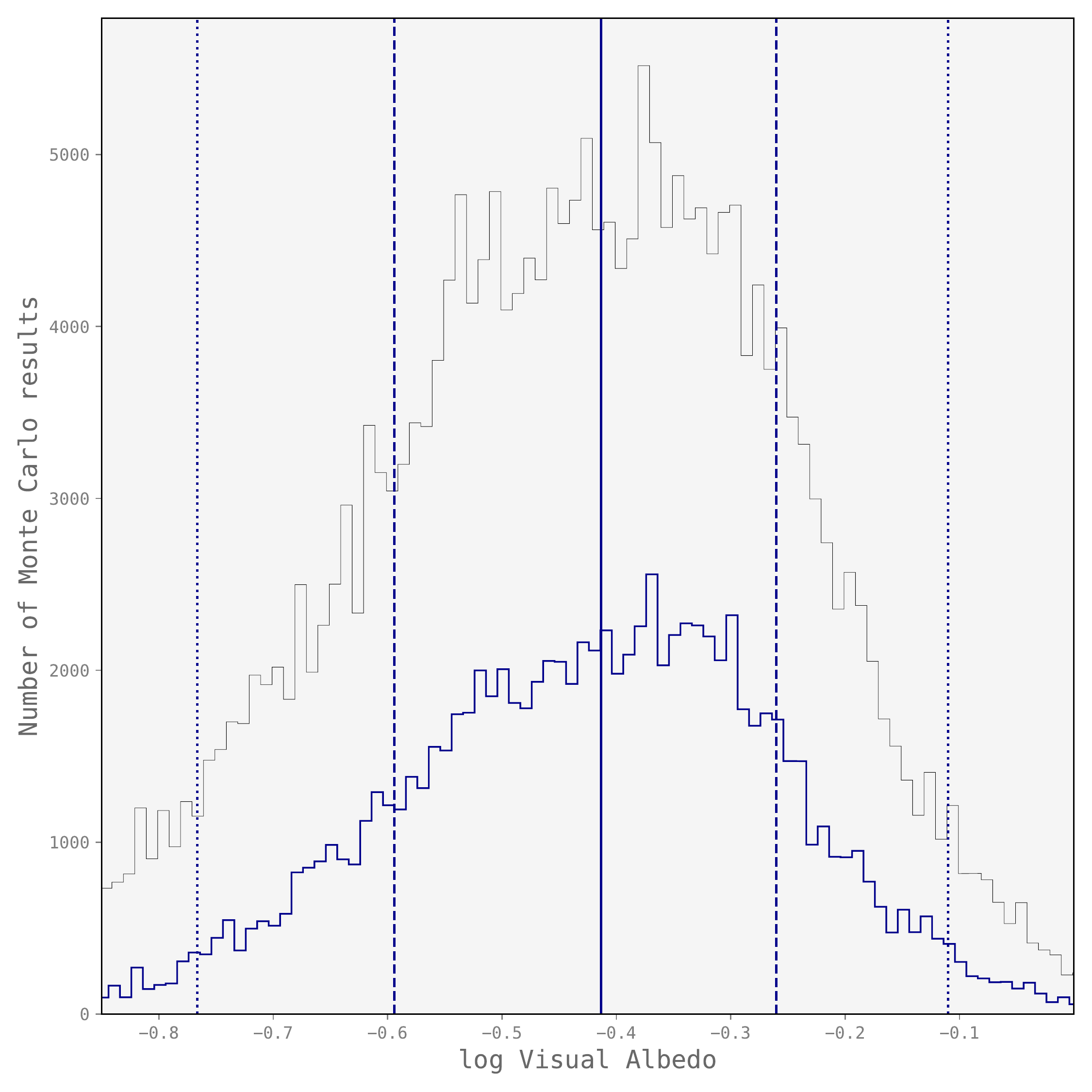}{0.5\textwidth}{}
          \fig{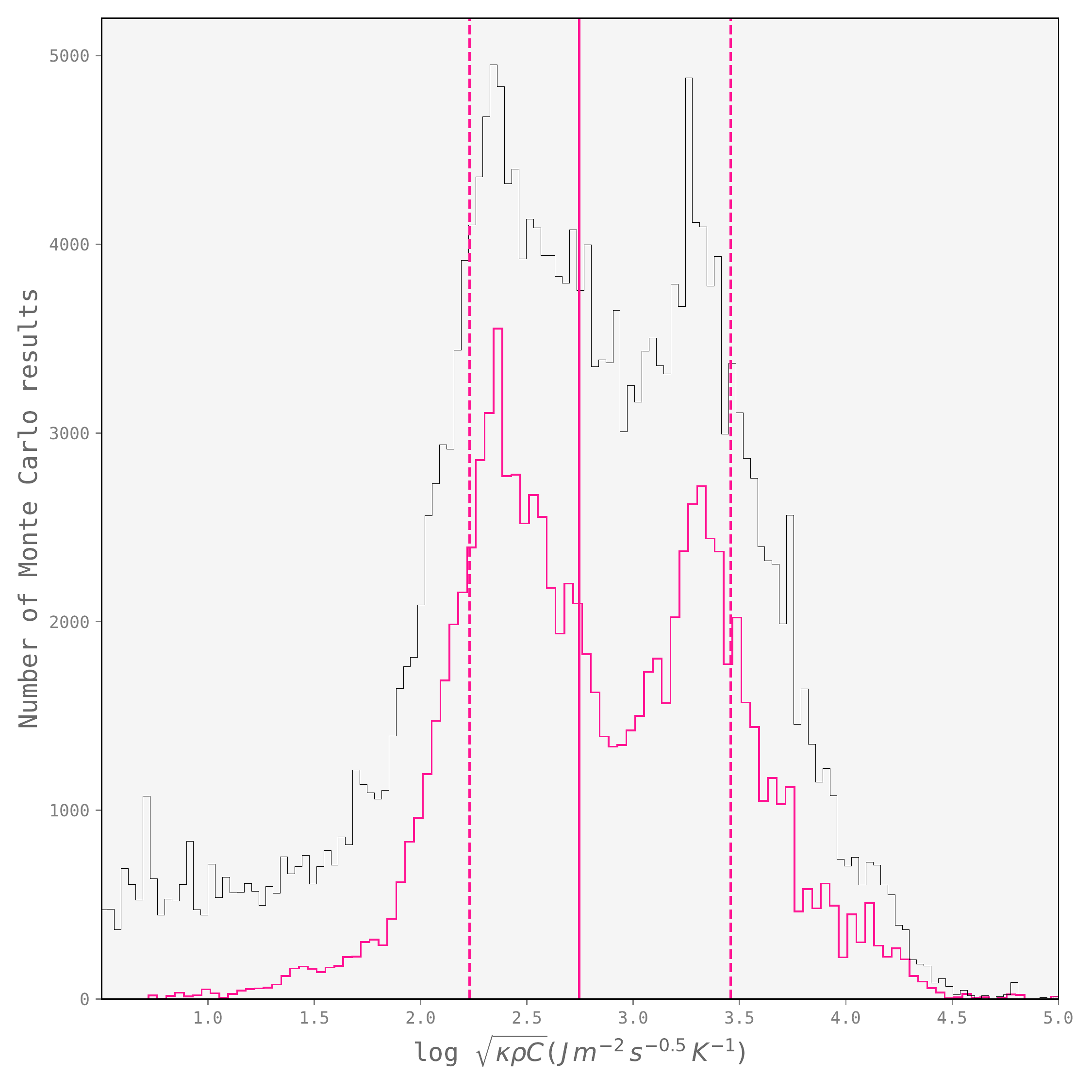}{0.5\textwidth}{}
          }
\caption{Plots highlighting the TPM results with (99942) Apophis modeled as a triaxial ellipsoid. In a clockwise manner from the top-left: average flux density during the two epochs along with the span of the bandpasses (the average flux was 95.37 {\textmu}Jy in W1 and 385.73 {\textmu}Jy in W2 during the first epoch, and 621.31 {\textmu}Jy in W1 and 3496.75 {\textmu}Jy in W2 during the second epoch), histogram of the diameter results with the 68\% (dashed lines) and 95\% (dotted lines) confidence interval ranges, histogram of the thermal inertia results with the 68\% (dashed lines) confidence interval range (the 95\% confidence interval values lie outside the x-axis range), and histogram of the visual albedo results with the 68\% (dashed lines) and 95\% (dotted lines) confidence interval ranges. The fainter histograms in the top-right, bottom-left, and bottom-right plots show the distribution without the LCDB pole solution constraint.\label{fig:TPMtriplots}}
\end{figure*}

\section{\textbf{Thermal Models and Results}} \label{sec:MandR}

Categorizing asteroids based on diameter and albedo by modeling their thermal emission (dominant in wavelengths greater than 4 µm for most near-Earth objects and asteroids closer than $\sim$4 au heliocentric distance) has become standard practice since infrared observations of small bodies began in the 1970s. Standard thermal models idealize asteroids as non-rotating spheres with a zero-degree solar phase angle. The asteroid model is bisected into a day and a night side, and it is assumed that the thermal emission on the dayside decreases from the subsolar point to the terminator, and that the night side has zero thermal emission. Following the temperature distribution, the reflected and absorbed sunlight on the dayside of the simplified asteroid model can be considered to be in equilibrium and can be used to compute the effective spherical diameter and bolometric Bond albedo. Next, the V-band geometric albedo can be determined using the system laid out by Bowell et al. \citeyearpar{bowell_89}, which uses the absolute magnitude (H) and phase slope coefficient (G) to describe the phase curve function. Additional parameters, including the so-called beaming parameter, can be used to better constrain the results by correcting for variation in thermal emission due to surface and shape irregularities, thermal inertia and rotation, spectral slope, and differing observing geometries (see Matson \citeyear{matson_b}, Morrison \citeyear{morrison_73}, Jones \& Morrison \citeyear{jones_morrison_94}, Morrison \& Chapman \citeyear{morrison_chapman_76}, Morrison \& Lebofsky \citeyear{morrison_lebof_79}, Lebofsky \& Spencer \citeyear{Lebofsky_spencer_89}, and Harris \& Lagerros \citeyear{harris_2002} for an extended discussion on the topic).\\

Lebofsky et al. \citeyearpar{Lebofsky_stm} computed an improved estimate of the beaming parameter based on observations of 1 Ceres and 2 Pallas and put forward the refined Standard Thermal Model (STM). While this model was demonstrated to effectively determine the size and visual albedo of several MBAs, the same estimations for NEAs had significant deviations when compared to radiometric predictions (Veeder et al. \citeyear{veeder_89}). Unlike the two very large main-belt objects that were used to calibrate the parameters of the refined STM, a typical NEA is observed at higher phase angles and has distinct physical and thermal properties. These confounding factors likely decreased the fidelity of the refined STM fits for the NEAs.\\

Harris \citeyearpar{Harris_neatm} developed a modified version of the refined STM called the Near-Earth Asteroid Thermal Model (NEATM) and presented revised diameter and visual albedo estimates with substantially better matches to independently measured diameters for several NEAs. Two primary improvements were implemented by \citeauthor{Harris_neatm}: first, incorporating observing geometry in the calculations to account for flux variations that arise on the nightside and the dayside; and second, including a varying beaming parameter to figure in the assumption that the nightside had zero emission and for other uncertainties due to shape, spin pole, rotation rate, surface roughness, and thermal conductivity. The NEATM is capable of fitting the diameter, beaming, visible geometric albedo, and under certain circumstances, the albedo at shorter infrared wavelengths (3 - 4 {\textmu}m). The precise number of parameters that can be fit depends on the number of measurements available. For asteroids detected by the reactivated NEOWISE mission, the brightnesses at 3.4 {\textmu}m and/or 4.6 {\textmu}m, and (in most cases) the visible magnitude are available from archival observations. Follow-up studies such as Mainzer et al. \citeyearpar{mainzer_11_neatm_proof} and Wright et al. \citeyearpar{response_to_mhyrvold} have employed NEATM and confirmed the reliability of the thermal model.\\

To compute physical characteristics of asteroids beyond the size and albedo, complex models called thermophysical models have been formulated by Lagerros \citeyearpar{Lagerros_96}, Rozitis \& Green \citeyearpar{rozitis_TPM}, {Hanu{\v{s}}} et al. \citeyearpar{Hanus_TPM}, and others. We used the Spherical, Cratered, Rotating, Energy-conserving Asteroid Model (simply called TPM in this paper) --- developed by Wright \citeyearpar{Wright_TPM} and Koren et al. \citeyearpar{Koren_TPM} and validated by Masiero et al. \citeyearpar{joe_TPM} --- which models the asteroid as a rotating cratered sphere (or a tri-axial ellipsoid) and uses thermal IR flux measurements to determine its diameter, albedo, and thermal inertia. Unlike STM or NEATM where surface roughness is approximated by the beaming parameter, the TPM varies the fraction of the craters on the facets, and calculates the thermal emission of all the idealized craters by accounting for the incident solar flux, blackbody radiation, solar reflection, and heat conduction. Best fits for up to ten different free parameters --- the RA and Dec of the spin axis pole position, diameter, visual albedo, rotational period, thermal inertia, cratering fraction, the p\textsubscript{IR}/p\textsubscript{V} ratio, and the b/a and c/b axis ratios --- can be computed by employing an affine-invariant Markov Chain Monte Carlo simulation (adapted from Foreman-Mackey et al. \citeyear{emcee}).\\

Priors for the different parameters mentioned above are assumed. For rotational pole position, the prior is uniform in 4$\pi$. The prior for the diameter is a log-uniform distribution of values between 1 m - 1000 km, and the same for visual albedo is a mixture model of two Rayleigh distributions given by Wright et al. \citeyearpar{wright_17}. The rotational period prior is modeled as a log-Cauchy distribution in terms of the equatorial rotational velocity but is heavily penalized for periods $\leq$ 2 hours if D $\geq$ 200 m (although for Apophis, the rotational period was fixed as described in the last paragraph of this section). The prior for the cratering fraction ($f_c$) is uniform in 0 - 1, and a penalty of $-2\,ln[f_c\,(1\,-\,f_c)\,/\,4]$ is added to the $\rchi^2$ to make the prior uniform in $f_c$. The surface roughness is parametrized by the cratering fraction with a slope of 75\textsuperscript{${\circ}$} and a variable fraction of flat terrain. The prior of the thermal inertia ($\Gamma$) is a log-uniform distribution for values 2.5 - 2500 Jm{\textsuperscript{-2}}s{\textsuperscript{-{\onehalf}}}K{\textsuperscript{-1}} with a width of one unit of natural log. And finally, the prior of p\textsubscript{IR}/p\textsubscript{V} is 0.563 $\pm$ 0.340 based on Mainzer et al. \citeyearpar{mainzer_prelimresults_11}.\footnote{In the triaxial ellipsoid shape model, along with the eight parameters previously listed, uniform priors are assumed for (b/a)\textsuperscript{4} and (c/b)\textsuperscript{4} in the range 0 - 1. The distribution of b/a is supplied by the lightcurve amplitudes, and the diameter is computed as $D = 2*(abc)^{\frac{1}{3}}$.} Cutoffs are enforced for the period and $\Gamma$ prior by adding a penalty to the $\rchi^2$ when a model is outside the allowed range. These penalties are of the form $[ln(parameter/limit)/width]^2$ but only applied if parameter is greater than the upper limit or if the parameter is less than the lower limit. The walkers update their position and compute a new parameter set using the equation $p_t\,=\,p_1\,+\,(p_2\,-\,p_1)\,z$, where $z$ is in the range 0.5 to 2 and the square root of $z$ is uniformly distributed, and ultimately, the goodness of fit is determined by a robust chi-square test.\\

For (99942) Apophis, measurements at both 3.4 {\textmu}m and 4.6 {\textmu}m bands were applied to the NEATM and TPM to derive size and visual albedo estimates. Thermal inertia ($\Gamma$), crater fraction, and axis ratios were also calculated by the TPM.\\

For implementing NEATM, values for the beaming parameter and the ratio of infrared to visible albedos (p\textsubscript{V}/p\textsubscript{IR}) were assumed based on prior measurements of these quantities from cases where more thermally-dominated infrared bands were available. Beaming was assumed to be 1.4 $\pm$ 0.5 and p\textsubscript{W1}/p\textsubscript{V} was assumed to be 1.6 $\pm$ 1.0 based on Mainzer et al. \citeyearpar{mainzer_w1factor} and Masiero et al. \citeyearpar{neowise_yr6and7}; the slope parameter G was set to 0.25 $\pm$ 0.2 --- a value appropriate for S-type asteroids (Delbo et al. \citeyear{delbo} and Vere{\v{s}} et al. \citeyear{veres_panstarrs_G}) --- and the H-mag was set to the derived value of 19.1 $\pm$ 0.5. For the first epoch, only the W2 band data were used as observations in W1 were too faint to produce reliable results. However, the object was much brighter during the second epoch of observations, and data from both the bands were used. NEATM analysis on the first epoch data returned a diameter of 306 $\pm$ 86 m and a geometric albedo of 0.43 $\pm$ 0.24, and applying the same on the second epoch data yielded a diameter of 406 $\pm$ 123 m and a visible geometric albedo of 0.29 $\pm$ 0.20. The average effective spherical diameter was calculated to be 355 $\pm$ 75 m, and the average visible geometric albedo was found to be 0.36 $\pm$ 0.16. Results obtained were in agreement with systematic uncertainties associated with NEOWISE diameter and albedo estimates derived solely using 3.4 {\textmu}m and 4.6 {\textmu}m photometry (expected to be $\sim$20\% and $\sim$40\%, respectively, as shown by Mainzer et al. \citeyear{mainzer_2012_760} and Masiero et al. \citeyear{masiero_12}). Figure \ref{fig:lightcurves} contains the lightcurve information from the two sets of observations.\\ 

For the TPM, the average flux was computed for each band at each observing epoch. First, the spherical shape model (also denoted as spherical TPM in this paper) was employed with a fixed rotational period of 30.568 hours,\footnote{The rotational period was computed by averaging the five different results listed in Loera-Gonz{\'a}lez et al. \citeyearpar{rotation_period} along with the rotational period derived by Augustin and Behrend \href{https://obswww.unige.ch/~behrend/page5cou.html\#099942}{(2021)}.} a G-value of 0.25, an H-value of 19.1 $\pm$ 0.5, and an emissivity of 0.9. Two hundred priors were generated, and after 48,600 Markov chain loops, the TPM yielded an effective diameter of 340 $\pm$ 65 m, an effective visible geometric albedo of 0.31 $\pm$ 0.09, and thermal inertia ($\Gamma$) in the range of 350 - 3400 Jm{\textsuperscript{-2}}s{\textsuperscript{-{\onehalf}}}K{\textsuperscript{-1}} with a best fit value of 950 Jm{\textsuperscript{-2}}s{\textsuperscript{-{\onehalf}}}K{\textsuperscript{-1}}. The crater fraction was found to be $0.49^{+0.32}_{-0.31}$. The thermophysical results were further constrained using the LCDB (Warner et al. \citeyear{PDS_data_system}) pole solution for Apophis --- ($\alpha$, $\delta$) = (119\textsuperscript{${\circ}$}, -79\textsuperscript{${\circ}$}) with a search  radius of 50\textsuperscript{$\circ$}. In this case, the TPM yielded an effective diameter of 330 $\pm$ 50 m, an effective visible geometric albedo of 0.32 $\pm$ 0.08, $\Gamma$ $\sim$ 300 - 2150 Jm{\textsuperscript{-2}}s{\textsuperscript{-{\onehalf}}}K{\textsuperscript{-1}} with a best fit value of 700 Jm{\textsuperscript{-2}}s{\textsuperscript{-{\onehalf}}}K{\textsuperscript{-1}}, and a crater fraction equal to $0.46^{+0.33}_{-0.28}$.\\

(99942) Apophis was also modeled as a triaxial ellipsoid in the TPM (referred to as triaxial TPM in this paper) and run with the same rotational period, H, G, and emissivity values as the spherical model. With the triaxial implementation, the TPM yielded an effective diameter of 330 $\pm$ 90 m, an effective visible geometric albedo of 0.31 $\pm$ 0.10, a crater fraction of $0.51^{+0.34}_{-0.34}$, b/a axis ratio equal to $0.63^{+0.10}_{-0.12}$, c/b axis ratio equal to $0.85^{+0.11}_{-0.21}$, and thermal inertia ($\Gamma$) in the range of 100 - 2800 Jm{\textsuperscript{-2}}s{\textsuperscript{-{\onehalf}}}K{\textsuperscript{-1}} with a best fit value of 500 Jm{\textsuperscript{-2}}s{\textsuperscript{-{\onehalf}}}K{\textsuperscript{-1}}. Similar to the spherical model, the triaxial TPM implementation was further constrained using the LCDB pole solution for Apophis. This yielded an effective diameter of 340 $\pm$ 70 m, an effective visible geometric albedo of 0.31 $\pm$ 0.09, a crater fraction of $0.52^{+0.34}_{-0.34}$, b/a axis ratio equal to $0.63^{+0.10}_{-0.15}$, c/b axis ratio equal to $0.87^{+0.10}_{-0.18}$, and $\Gamma$ $\sim$ 150 - 2850 Jm{\textsuperscript{-2}}s{\textsuperscript{-{\onehalf}}}K{\textsuperscript{-1}} with a best fit value of 550 Jm{\textsuperscript{-2}}s{\textsuperscript{-{\onehalf}}}K{\textsuperscript{-1}}. The plots in figure \ref{fig:TPMplots} and figure \ref{fig:TPMtriplots} are associated with the TPM, and they display the flux during the two epochs, the effective spherical diameter of the object, the best fit thermal inertia value, and the distribution of the visual albedo results.\\

\section{\textbf{Discussion}} \label{sec:discussion}

Delbo et al. \citeyearpar{delbo} were the first to present size and visible albedo estimations of (99942) Apophis using polarimetric observations from the 8.2 m VLT-Kueyen telescope of the European Southern Observatory, finding p\textsubscript{V} = 0.33 $\pm$ 0.08 and D\textsubscript{eff} = 270 $\pm$ 60 meters. Müeller et al. \citeyearpar{mueller} used far-infrared observations from the Herschel Space Telescope PACs instrument and estimated D\textsubscript{eff} = $375^{+14}_{-10}$ m and p\textsubscript{V} = $0.3^{+0.05}_{-0.06}$. Licandro et al. \citeyearpar{GTC_thermalinertia} combined Herschel-PACS data with GTC/CanariCam data and reported an effective diameter between 380 - 393 meters and a visible albedo in the range 0.24 - 0.33. {Brozovi\'{c}} et al. \citeyearpar{brozovic} found a best fit diameter of 340 $\pm$ 40 m and a visible albedo of 0.35 $\pm$ 0.10 by examining (99942) Apophis using radar observations from Goldstone and Arecibo, and rendering it as a 3D model using lightcurve-derived shape and spin states from Pravec et al. \citeyearpar{pravec_Hmag}. Our effective diameter and visible albedo estimate for Apophis from both NEATM and TPM are consistent with estimates put forward by Müeller et al. \citeyearpar{mueller}, Licandro et al. \citeyearpar{GTC_thermalinertia}, and {Brozovi\'{c}} et al. \citeyearpar{brozovic} to within the measurement uncertainties.\\

Our results show that Apophis most likely has a thermal inertia value $\sim$ 550 Jm{\textsuperscript{-2}}s{\textsuperscript{-{\onehalf}}}K{\textsuperscript{-1}, and such a value suggests that the asteroid’s surface may consist of small and moderately-sized boulders possibly interspersed with coarse sand.} Our derived thermal inertia result overlaps with estimated ranges cited in the previous studies. The same study by Müeller et al. \citeyearpar{mueller} found a thermal inertia in the range of 250 - 800 Jm{\textsuperscript{-2}}s{\textsuperscript{-{\onehalf}}}K{\textsuperscript{-1}} with a best fit at 600 Jm{\textsuperscript{-2}}s{\textsuperscript{-{\onehalf}}}K{\textsuperscript{-1}}, and Licandro et al. \citeyearpar{GTC_thermalinertia} cited a thermal inertia in the range of 50 - 500 Jm{\textsuperscript{-2}}s{\textsuperscript{-{\onehalf}}}K{\textsuperscript{-1}}. A much lower value of $\Gamma = 100^{+240}_{-100}$ Jm{\textsuperscript{-2}}s{\textsuperscript{-{\onehalf}}}K{\textsuperscript{-1}} was also put forward by Yu et al. \citeyearpar{yu}. For comparison, Itokawa --- an asteroid very similar to Apophis in size and taxonomic class --- was measured to have a $\Gamma = 700 \pm 200$ Jm{\textsuperscript{-2}}s{\textsuperscript{-{\onehalf}}}K{\textsuperscript{-1}} by Müeller et al. \citeyearpar{itokawa_TI}. However, low thermal inertia values cannot be dismissed given the case of Bennu ($\Gamma$ $\sim$ 300 Jm{\textsuperscript{-2}}s{\textsuperscript{-{\onehalf}}}K{\textsuperscript{-1}}) where surface property predictions based on remote observations were incongruent with those based on in-situ observations (Rozitis et al. \citeyear{Bennu_thermal}).\\

The TPM may seem preferable over the NEATM since it can be used to compute parameters beyond diameter and albedo such as spin axis or thermal inertia; however, both thermal models are useful depending on the type and quantity of the data, and the goal of the study. The NEATM is highly computationally efficient, consuming seconds to minutes to produce results for a single object such as Apophis, whereas the TPM takes several hours to do the same. The NEATM is most appropriate for objects where only a single viewing geometry and a limited number of wavelengths are available. On the other hand, the TPM is suitable in cases where multi-epoch observations in a thermally-dominated wavelength are available. In the case of (99942) Apophis, both epochs of W2 observations are thermally dominated as the heliocentric distance of the object was roughly 1 au.\\

\begin{figure*}
\plotone{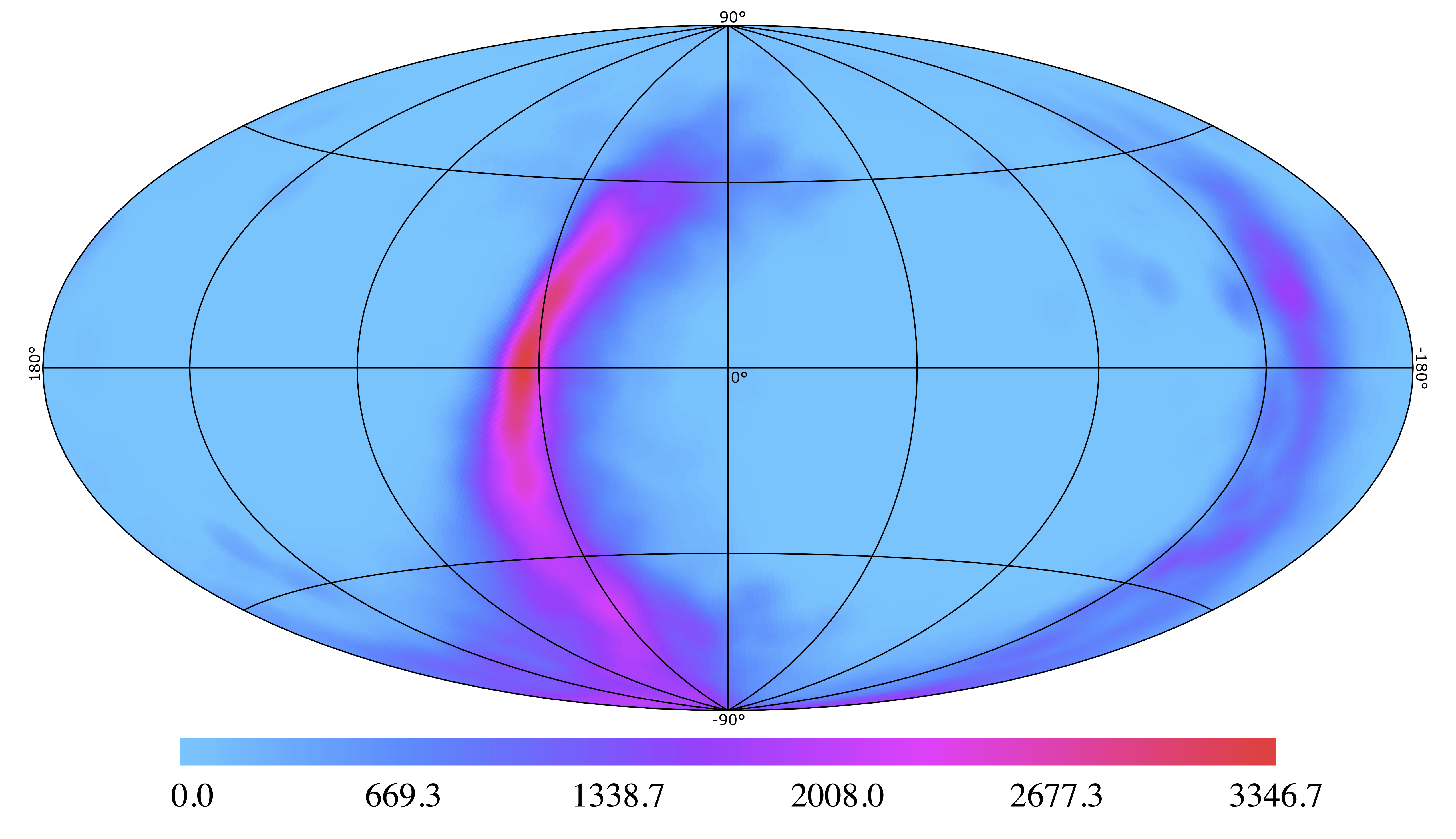}
\caption{Spin pole position solution with emissivity of 0.9 as constrained by the triaxial TPM plotted in celestial coordinates. The peak pole position was found at an RA of 50.6\textsuperscript{$\circ$} and a Dec of 0\textsuperscript{$\circ$}. The color bar denotes the number of solutions in the bin.
\label{fig:pole_pos}}
\end{figure*}

Regardless of the thermal model used, appropriate selection of the phase curve slope parameter G is necessary to accurately determine the apparent visible magnitude from the absolute magnitude (H) at the time of the NEOWISE IR observations. For an S-class or an Sq-class asteroid like Apophis, the G-value is typically assumed to be 0.25. We used the average phase integral value (\textit{q} = 0.47 $\pm$ 0.15) put forward by Shevchenko et al. \citeyearpar{shevchenko_phase} to calculate a G = 0.25 $\pm$ 0.20. Veres et al. \citeyearpar{veres_panstarrs_G} and Colazo et al. \citeyearpar{Colazo_2021} have computed a value of G $\sim$ 0.20 for such asteroids, but this value is within the measurement uncertainty range and does not seem to have a noticeable effect on the results. For instance, taking G = 0.2 $\pm$ 0.1 gives a result of D\textsubscript{eff} = 355 $\pm$ 79 m and p\textsubscript{V} = 0.387 $\pm$ 0.302 using the NEATM, and a D\textsubscript{eff} = 365 $\pm$ 63 m and p\textsubscript{V} = 0.33 $\pm$ 0.10 via the TPM; while larger variations in the slope parameter would affect the resultant visual albedo, estimates for G = 0.20 $\pm$ 0.10 and G = 0.25 $\pm$ 0.10 are virtually identical. Small changes in the absolute magnitude H, however, drastically affect V-albedo as shown by Masiero et al. \citeyearpar{masiero_HV_21}. The Horizons catalog H-mag of 19.7 was, thus, not used in the thermal models because H estimates for NEAs tend to be imprecise due to poor sampling of the phase curve over a small range of phase angles. Near-simultaneous photometry from CTIO enabled us to calculate H = 19.1 $\pm$ 0.5 (for comparison, Pravec et al. \citeyear{pravec_Hmag} reported a similar H-value of 19.09 $\pm$ 0.19).\\ 

Considerations regarding the choice of emissivity value and pole positions were also made. Studies show that meteorite samples and asteroid analogs typically have emissivities close to 0.9 in the low-IR and mid-IR ranges (Donaldson Hanna et al. \citeyear{hanna_orex_em} and Maturilli et al. \citeyear{maturilli_analogs_em}). Rozitis et al. \citeyearpar{rozitis_em} noted that an emissivity of 0.9 is a fair assumption at all observed wavelengths as thermophysical models have been able to reproduce 4 - 40 {\textmu}m observations of various asteroids. We quantified the influence of emissivity by running the TPM with the NEOWISE's Apophis observations using emissivity values of 0.8, 0.9, 0.95, and 0.99. No significant differences were observed in the TPM results for the four different emissivity values (see Table \ref{tab:apophis_em_table}). The mean pole positions computed in the four scenarios were also virtually identical (spin-pole results for emissivity of 0.9 can be found in Figure \ref{fig:pole_pos}). Thus, an emissivity of 0.9 was used for the final TPM implementation.\\

\tabcolsep=0.1cm
\begin{deluxetable}{cccc}
\tablenum{4}
\tablecaption{TPM results for (99942) Apophis for four different emissivity values.\label{tab:apophis_em_table}}
\tablewidth{0pt}
\tablehead{
\colhead{Emissivity} & \colhead{Diameter} & \colhead{Albedo} & \colhead{Thermal inertia}\\
\nocolhead{} & \colhead{\footnotesize{Meters}} & \nocolhead{} & \colhead{\footnotesize{Jm{\textsuperscript{-2}}s{\textsuperscript{-{\onehalf}}}K{\textsuperscript{-1}}}} 
}
\startdata
0.80 & 320 $\pm$ 60 & 0.33 $\pm$ 0.10 & $1050^{+2150}_{-700}$ \\
0.90 & 340 $\pm$ 65 & 0.31 $\pm$ 0.09 & $950^{+2450}_{-600}$ \\
0.95 & 340 $\pm$ 70 & 0.31 $\pm$ 0.09 & $900^{+2750}_{-600}$ \\
0.99 & 350 $\pm$ 70 & 0.31 $\pm$ 0.09 & $900^{+2400}_{-600}$ \\
\tableline
0.80 & 320 $\pm$ 80 & 0.31 $\pm$ 0.10 & $400^{+2250}_{-350}$ \\
0.90 & 330 $\pm$ 90 & 0.30 $\pm$ 0.10 & $500^{+2300}_{-400}$ \\
0.95 & 330 $\pm$ 80 & 0.31 $\pm$ 0.10 & $400^{+1850}_{-350}$ \\
0.99 & 330 $\pm$ 90 & 0.30 $\pm$ 0.10 & $350^{+1800}_{-300}$ \\
\enddata
\tablecomments{The results shown are TPM implementations without the LCDB pole solution constraint. The top four rows are results from the spherical TPM, and the bottom four are from the triaxial TPM.\label{fig:tpm_diff_em_results}}
\end{deluxetable}

Reflected light contributions were also considered to ensure that the bands were thermally dominated during the observations. In our implementation of the NEATM, if more than three-quarters of the light in the first band comprises of reflected light, then fitting is redone with the infrared albedo as a free parameter (the thresholds are different if three or more detections exist). In the case of NEOWISE's (99942) Apophis observations, the reflected light contributions were $\sim$57\% in W1 and $\sim$1\% in W2 in Epoch 2. The same for the latter was $\sim$2\% in Epoch 1 (W1 observations from Epoch 1 were discarded during the NEATM fitting routine due to their low signal to noise ratios.). Therefore, W2 was thermally dominated during both the observing epochs, indicating that there is sufficient thermal flux for thermal modeling.\\

In this paper, the TPM was employed using both the LCDB pole solution and the pole solution constrained using NEOWISE’s Apophis data. The use of the LCDB pole solution is intended to demonstrate how rotational state data, which could be obtained during a close-approach event, would improve the TPM results and greatly benefit Planetary Defense. The triaxial TPM was implemented under the assumption that if the object were to truly impact Earth, a coordinated campaign to determine the shape from light curve inversion would have been carried out. Although not part of the exercise as it was undertaken, such kind of data could have been acquired during a close-pass epoch if additional resources had been brought to bear. Further areas of improvement could include combining NEOWISE data with Herschel’s and implementing the triaxial (or a more complex) shape model in the TPM to better constrain the size, albedo, and thermal inertia. The same could be achieved through multi-epoch thermal observations as well, although such observations are hard to obtain from NEOWISE due to a fixed scanning pattern and limited mission time. Upcoming projects like the Near-Earth Object Surveyor (NEOS; Mainzer et al. \citeyear{neo_surveyor}) could make it possible to obtain multi-epoch data in wavelengths dominated by the thermal emission of asteroids. Future studies could theorize the effect of a large thermal inertia value on the Yarkovsky drift (Bottke et al. \citeyear{yarkovsky}) and close-approach gravitational perturbations (Chodas \citeyear{chodas_keyholes}).

\tabcolsep=0.1cm
\begin{deluxetable*}{lccc}
\tablenum{5}
\tablecaption{Table summarizing the thermophysical results for (99942) Apophis.\label{tab:apophis_res}}
\tablewidth{0pt}
\tablehead{
\colhead{Diameter} & \colhead{Albedo} & \colhead{Thermal inertia} & \colhead{Reference}\\
\colhead{\footnotesize{Meters}} & \nocolhead{} & \colhead{\footnotesize{Jm{\textsuperscript{-2}}s{\textsuperscript{-{\onehalf}}}K{\textsuperscript{-1}}}} & \nocolhead{} 
}
\startdata
270 $\pm$ 60 & 0.33 $\pm$ 0.08 & - & Delbo et al. \citeyearpar{delbo} \\
$375^{+14}_{-10}$ & $0.30^{+0.05}_{-0.06}$ & - & Müeller et al. \citeyearpar{mueller} \\
380 - 393 & 0.24 - 0.33 & 50 - 500 & Licandro et al. \citeyearpar{GTC_thermalinertia}\\
340 $\pm$ 40 & 0.35 $\pm$ 0.10 & 600 & {Brozovi\'{c}} et al. \citeyearpar{brozovic}\\
340 $\pm$ 70 & 0.31 $\pm$ 0.09 & 550 & This Paper\\
\enddata
\end{deluxetable*}

\section{\textbf{Conclusion}} \label{sec:conclusion}

Independent measurements from NEOWISE helped estimate (99942) Apophis’ size, visual albedo, and thermal inertia during the 2020-21 planetary defense exercise. As a summary, our analyses returned the following results:

\begin{itemize}
  \item An absolute visual magnitude of 19.1 $\pm$ 0.5 using a slope parameter of 0.25 $\pm$ 0.20.
  \item From the NEATM, an effective spherical diameter of 355 $\pm$ 75 m and an average geometric albedo of 0.36 $\pm$ 0.16.
  \item From the spherical TPM with the LCDB pole solution, an effective diameter of 330 $\pm$ 50 m, an effective geometric albedo of 0.32 $\pm$ 0.08, and a $\Gamma$ = $700^{+1450}_{-400}$
  Jm{\textsuperscript{-2}}s{\textsuperscript{-{\onehalf}}}K{\textsuperscript{-1}}.
  \item From the triaxial TPM with the LCDB pole solution, an effective diameter of 340 $\pm$ 70 m, an effective geometric albedo of 0.31 $\pm$ 0.09, and a $\Gamma$ = $550^{+2300}_{-400}$
  Jm{\textsuperscript{-2}}s{\textsuperscript{-{\onehalf}}}K{\textsuperscript{-1}}.
\end{itemize}

The estimated diameter derived from the ``discovery" observations allowed exercise participants to calculate that the object would produce $\sim$8.5$\times$10\textsuperscript{19} J (or 2$\times$10\textsuperscript{7} kilotons of TNT) of energy upon impact.\footnote{Density was assumed to be 2700 kgm\textsuperscript{-3} from Carry \citeyearpar{carry_density}, and impact velocity was assumed to be 20 kms\textsuperscript{-1} from Harris \& Hughes \citeyearpar{ast_collision_v} and French \citeyearpar{catastrophe_v}.} Although this estimate is a factor of 10\textsuperscript{4} larger than the estimated energy produced by Chelyabinsk, the participants quickly determined that the object would most likely cause a regional disaster and not a global one.\\

Thus, we demonstrate that the NEOWISE data and thermal modeling are capable of rapid and accurate physical characterization of Near-Earth Asteroids (NEAs), and highlight that such expeditious data processing could prove to be essential in assessing threat levels and mitigating hazards posed by them.\\

\section{\textbf{Acknowledgements}}

The NEOWISE Reactivation Mission is funded by the NASA Science Mission Directorate Planetary Science Division as part of the Planetary Defense Coordination Office. Dr. Amy Mainzer of the University of Arizona is the NEOWISE Principal Investigator. NEOWISE is managed and operated by JPL. Data processing, archiving and distribution for NEOWISE is carried out by IPAC, California Institute of Technology. Operations of the NEOWISE spacecraft and payload are supported by Ball Aerospace and Technology Corp. and the Space Dynamics Laboratory, Utah State University.\\

NEOWISE gratefully acknowledges the services contributed by the IAU Minor Planet Center, operated by the Harvard-Smithsonian Center for Astrophysics.\\

NEOWISE makes use of data from the WISE, AllWISE, and original NEOWISE projects that were funded by the NASA Astrophysics and Planetary Science Divisions.\\

The work of EK and JP was conducted at the Jet Propulsion laboratory, California Institute of Technology, under a contract with the National Aeronautics and Space Administration (80NM0018D0004).\\

We are also thankful for the High Performance Computing (HPC) resources provided by the University of Arizona TRIF, UITS, and Research, Innovation, and Impact (RII) and maintained by the UArizona Research Technologies department.\\

\software{astropy (Astropy Collboration et al. \citeyear{astropy1}, \citeyear{astropy2}), scipy (Virtanen et al. \citeyear{scipy}), numpy (Harris et al. \citeyear{numpy})}


\bibliography{1-preprint}{}
\bibliographystyle{aasjournal}

\end{document}